\magnification=\magstephalf
\input amstex
\loadbold
\documentstyle{amsppt}
\refstyle{A}
\NoBlackBoxes

\vsize=7.5in

\def\pf{\hfill $\square$}
\def\c{\cite}

\def\fg{\frak{g}}

\def\end{\text{End}}

\def\bl{\boldkey l}

\def\bk{\boldkey K}
\def\bl{\boldkey L}
\def\ba{\boldkey A}
\def\bb{\boldkey B}

\def\bi{\boldkey I}
\def\bj{\boldkey J}

\topmatter
\title Long time behaviour for a class of low-regularity 
solutions of the Camassa-Holm equation\endtitle
\leftheadtext{L.-C. Li}
\rightheadtext{Camassa-Holm Equation}

\author Luen-Chau Li\endauthor
\address{L.-C. Li, Department of Mathematics,Pennsylvania State University,
University Park, PA  16802, USA}\endaddress
\email luenli\@math.psu.edu\endemail

\abstract  In this paper, we investigate the long time behaviour for
a class of low-regularity solutions of the Camasssa-Holm equation given
by the superposition of infinitely many interacting traveling waves with
corners at their peaks.
\endabstract
\endtopmatter

\document
\subhead
1. \ Introduction
\endsubhead

\baselineskip 15pt
\bigskip

The Camassa-Holm (CH) equation 
$$u_t + 2\kappa u_x -u_{xxt} + 3uu_x = 2u_{x}u_{xx} + uu_{xxx}\eqno(1.1)$$
was derived in \c{CH} as a model of long waves in shallow water.
Since then, it has received considerable attention.
In this work, we will be concerned with the non-dispersive case on
the line, corresponding to $\kappa =0,$ that is,
$$u_t -u_{xxt} + 3uu_x = 2u_{x}u_{xx} + uu_{xxx},\quad x\in \Bbb R,\eqno(1.2)$$
and so from now on, we will refer to (1.2) as the CH equation.
It is a fundamental discovery of Camassa and Holm that (1.2) admits 
weak solutions, of the form
$$u(x,t) = c e^{-|x-ct|}\eqno(1.3)$$
for any nonzero constant $c.$  Clearly, any such solution is a
traveling wave with a corner at its crest or trough, depending
on the sign of $c.$ 
In \c{CH}, apparently motivated by such traveling wave solutions,
the authors proposed an even more general class of weak solutions of (1.2), 
namely,
$$u(x,t) = \frac{1}{2} \sum_{j=1}^{N} e^{-|x-q_{j}(t)|} p_{j}(t),\eqno(1.4)$$
where $N$ is a positive integer. Clearly, when $N >1$, this {\it ansatz \/}
represents the superposition of $N$ interacting (peak-shaped) waves
(called peakons if $p_j(t)>0$, antipeakons if $p_j(t)<0$)
in which case the $p_j$'s are no longer constants
and the $q_j$'s are no longer linear in $t.$
Indeed, the location of the peaks and their signed amplitudes
are now governed by the ordinary
differential equations \c{CHH}:
$$\eqalign{
&\dot{q}_{j} = \frac{1}{2} \sum_{k=1}^{N} e^{-|q_j -q_k|} p_k,\cr
&\dot{p}_{j} = \frac{1}{2} p_j \sum_{k=1}^{N} sgn(q_j -q_k)
   e^{-|q_j -q_k|} p_k,\quad j=1,\cdots, N\cr}\eqno(1.5)$$
which are the Hamiltonian equations of motion generated
by 
$$H = {1\over 4} \sum_{i,j=1}^{N} e^{-|q_i-q_j|} p_ip_j.\eqno(1.6)$$
For $N=2$, the equations in (1.5) were integrated
explicitly  and the long time behaviour was worked out in
\c{CHH}.   For $N >2,$  the explicit integration of
(1.5) in the sector where $q_1 <\cdots < q_N$ was obtained by 
Beals, Sattinger and Szmigielski
\c{BSS} using the inverse scattering method and
a theorem of Stieltjes on continued fractions \c{S}.
Furthermore, by making use of the explicit formulas
for the $q_j$'s and $p_j$'s, these authors were able
to analyze in detail the long time asymptotics of (1.5). 

In this work, we will consider a class of weak solutions of (1.2)
obtained by generalizing the ansatz in (1.4), namely, we
take
$$u(x,t) = \frac{1}{2} \sum_{j=1}^{\infty} e^{-|x-q_{j}(t)|} p_{j}(t),\eqno(1.7)$$
with $p_j(t) >0$ for all $j\in \Bbb N$ and such that
$p_j(t)\to 0$ sufficiently fast as $j\to \infty.$  More
precisely, we take 
$q(t) =(q_1(t),q_2(t),\cdots)\in l^{+}_{\infty},$
$p(t) =(p_1(t),p_2(t),\cdots)\in l^{+}_{1,2},$
where $l^{+}_{\infty}$ and $l^{+}_{1,2}$ are defined in (3.3).
Thus the equations of motion for $(q(t), p(t))$ are given by
$$\eqalign{
&\dot{q}_{j} = \frac{1}{2} \sum_{k=1}^{\infty} e^{-|q_j -q_k|} p_k,\cr
&\dot{p}_{j} = \frac{1}{2} p_j \sum_{k=1}^{\infty} sgn(q_j -q_k)
   e^{-|q_j -q_k|} p_k,\quad j\in \Bbb N.\cr}\eqno(1.8)$$
Our main goal in this work is to investigate the long time
behaviour for this class of solutions in the two sectors
$${\Cal S_{-}} = \lbrace (q,p)\in l^{+}_{\infty}\oplus l^{+}_{1,2} \mid
   q_1 < q_2 < \cdots, p_j > 0 \,\, \hbox{for all}\,\, j \rbrace,\eqno(1.9)$$
and 
$${\Cal S_{+}} = \lbrace (q,p)\in l^{+}_{\infty}\oplus l^{+}_{1,2} \mid
   q_1 > q_2 > \cdots, p_j > 0 \,\, \hbox{for all}\,\, j \rbrace.\eqno(1.10)$$
Once this is achieved, the adaptation of our analysis to other 
sectors defined by a restricted class of permutations of $\Bbb N$ is 
straightforward.
As the reader will see, our approach to this problem is based on the 
connection of (1.8) in the
sectors ${\Cal S}_{\pm}$ to the $(\pm)$ {\it Toda flow \/}
on some rather special Hilbert-Schmidt operators in 
$l^{+}_{2}$  whose kernels are parametrized by the $q_j$'s and the $p_j$'s.
  
The paper is organized as follows.  In Section 2, we begin by introducing
the Toda flows on Hilbert-Schmidt operators in $l^{+}_{2}$ using the
r-matrix approach, then we discuss the associated Hilbert Lie groups and the
factorization method to solve the so-called $(\pm)$ Toda flow.
We also quote from \c{DLT1} a preliminary result on the long time behaviour
of the $(\pm)$ Toda flow which is the point of departure for this work.
This section concludes with a study of the coadjoint orbits which are 
relevant in our studies.  In Section 3, we introduce the class of 
low-regularity solutions of the CH equation which we mentioned above 
and we establish the 
global existence of solutions of (1.8) in the sectors ${\Cal S}_{\pm}.$  Then
we make the connection with the $(\pm)$ Toda flow and we study
the spectral properties of the Lax operators which are basic
in our subsequent analysis.   In Section 4, we study the long
time behaviour of (1.7) and (1.8) in the sector ${\Cal S}_{-}.$  Here
the main difficulty is in showing that the peaked waves separate out, i.e.,
$\lim_{t\to \infty} |q_j(t) - q_k(t)| =\infty$ for $j\neq k.$
Technically, this is due to the fact that the (positive) eigenvalues
of the Lax operator $\bl(t)=\bl(q(t),p(t))$ accumulate at $0,$ which is 
in the essential spectrum of $\bl(t).$  The fact that $0$ is at the bottom
of the spectrum is also responsible for the result that
$\lim_{t\to\infty} p_j(t) =0$ for all $j\in \Bbb N.$  Finally,
we remark that although $q_j(t)$ still approaches $\infty$ as $t\to\infty,$
however, we only have $q_j(t) = o(t)$ in the present case.  Thus the
long time behaviour of (1.7) in the sector ${\Cal S}_{-}$ is quite
different from the analogous one in \c{BSS} for the multipeakon
solutions in (1.4).  In Section 5,
the final section, we first analyze the long time behaviour in the
sector ${\Cal S}_{+}.$   Here our main tool consists of the
induced derivations of the Lax operator, which are acting
on the $k$-th exterior powers $\wedge^{k} l^{+}_{2},\,\,\,k\geq 1.$
Note that this has been also used in \c{DLT1}, but in our
context we can express all the relevant quantities explicitly
in terms of eigenvalues
and eigenvectors.   An interesting feature here is that
the long time behaviour of the $(+)$ Toda flow $\bl(q(t),p(t))$
has the sorting property, as in the case of the Toda flow
on $N\times N$ Jacobi matrices \c{Mo}.  In particular, if
$0 < \cdots < \lambda_3 < \lambda_2 < \lambda_1$ are the
(simple) eigenvalues of $\bl(q(0),p(0))$, this means that
$\lim_{t\to\infty} p_j(t) = 2\lambda_j$ for all $j\in \Bbb N.$
In this case, the scattering behaviour follows easily
and we can show that $q_j(t)\sim \lambda_j t$ as $t\to \infty.$
Finally, motivated by a convincing but heuristic discussion in
[M], we end the paper by showing how our preceding analysis can be
adapted to other sectors of the phase space defined by a restricted 
class of permutations of $\Bbb N.$

\bigskip
\bigskip

\subhead
2. \ Toda flows on Hilbert-Schmidt operators and orbits of semiseparable 
\phantom{fak}\,\, operators
\endsubhead

\bigskip

Let  ${\Cal H}$ be the Hilbert space $l^{+}_{2}$ consisting of
sequences $u=(u_1, u_2,\cdots)$ of real numbers which satisfy
$\vert|u\vert| =\,(\sum_{i=0}^{\infty} u^{2}_{i})^{1\over 2} < \infty.$ 
In this section, we begin by introducing a class of isospectral
flows  on the space $\fg$ of Hilbert-Schmidt operators
on ${\Cal H}$  which is in some sense a natural generalization
of the Toda flows on $n\times n$ matrices \c{DLT2}.  Then
we will study some of the basic properties of these flows
which are relevant in our study of the Camassa-Holm equation.

Throughout the paper, let $B({\Cal H})$ be the space of bounded operators 
on ${\Cal H}.$  If $\ba\in B({\Cal H})$, we shall denote its transpose by
$\ba^{T}$, we shall also write $\ba = (A_{ij})_{i,j=1}^{\infty}$ if and only if
$$(\ba u)_{i} = \sum_{i=1}^{\infty} A_{ij}u_{j}, \quad i\geq 1,\,\, u\in l^{+}_{2}.
\eqno(2.1)$$
Thus  $(A_{ij})_{i,j=1}^{\infty}$ is the matrix representation
of $\ba$ with respect to the canonical basis $e_1,e_2,\cdots$ of $l^{+}_{2}$.
With this notation, we have the following basic fact, namely,
$\ba\in \fg$ if and only if the Hilbert-Schmidt norm
$\vert|\ba\vert|_{2}=\left(\sum_{i,j=1}^{\infty} A^{2}_{ij} \right)^{1\over 2} <\infty.$
Hence $\ba\in \fg$ implies $\ba^{T}$ is also in $\fg.$

We start with the following proposition which is an easy consequence of
the closure of $\fg$ under the operations of addition,
subtraction, and composition in $B({\Cal H}).$
 
\proclaim
{Proposition 2.1} $\fg$ is a Hilbert Lie algebra with Lie bracket
$[\cdot,\cdot]$ defined by
$$[\ba, \bb] = \ba\bb - \bb\ba \eqno(2.2)$$
and the inner product on $\fg$ is the usual Hilbert-Schmidt
inner product $(\cdot,\cdot)_{2}$, i.e.,
$$\eqalign{
(\ba, \bb)_{2}& = tr (\ba^{T}\bb)\cr
             & =\sum_{i,j=1}^{\infty} A_{ij}B_{ij}.\cr}\eqno(2.3)$$
Moreover, $\fg$ is equipped with the non-degenerate
ad-invariant pairing
$$(\ba, \bb)= \sum_{i,j=1}^{\infty} A_{ij}B_{ji}.\eqno(2.4)$$
\endproclaim

The Hilbert Lie algebra $\fg$ has two distinguished Lie subalgebras
$\frak l$ and $\frak k$, where $\frak l$ consists of lower triangular
operators $\ba\in \fg,$ i.e.,
$$A_{ij} = 0\,\,\, \hbox{for}\,\,\, i< j\eqno(2.5)$$
and $\frak k$ consists of  operators $\bb\in \fg$ for which
$$\bb^{T} = - \bb.\eqno(2.6)$$
We will call $\frak l$ the {\it lower triangular subalgebra} of $\fg$
and $\frak k$ the {\it skew-symmetric subalgebra}.   Our next
result is obvious. (See also (2.10) below.)

\proclaim
{Proposition 2.2} We have
$$\fg = \frak l \oplus \frak k.$$
\endproclaim

Let $\Pi_{\frak l}$ and $\Pi_{\frak k}$ be the projection operators
onto $\frak l$ and $\frak k$ respectively associated with
the splitting $\fg = \frak l\oplus \frak k.$  Then \c{STS}
$$ R = \Pi_{\frak l} - \Pi_{\frak k}\eqno(2.7)$$
is a classical r-matrix on $\fg$ satisfying the modified 
Yang-Baxter equation (mYBE) 
$$[R(\ba),R(\bb)] -R([R(\ba),\bb] + [\ba, R(\bb)]) = -[\ba,\bb]\eqno(2.8)$$
for all $\ba$, $\bb\in \fg.$  Consequently, the formula
$$[\ba, \bb]_{R} = {1\over 2}([R(\ba),\bb] + [\ba, R(\bb)]),\quad \ba,\bb\in
\fg \eqno(2.9)$$
defines a second Lie bracket on $\fg$  and we shall denote the associated
Lie algebra by  $\fg_{R}.$    Note that explicitly,
we have
$$\eqalign{
&\Pi_{\frak k} \ba = \ba_{+} - \ba^{T}_{+},\cr
&\Pi_{\frak l} \ba = \ba_{-} + \ba_{0} + \ba^{T}_{+}\cr}\eqno(2.10)$$
where for $\ba = (A_{ij})_{i,j=1}^{\infty}  \in \fg$, the operators 
$\ba_{+}$, $\ba_{-}$ and $\ba_{0}$ are defined respectively by the
strict upper triangular part, the strict lower triangular part
and the diagonal part of $(A_{ij})_{i,j=1}^{\infty}.$  In what follows, we 
will compute
the dual maps of all linear operators on $\fg$ with respect
to the pairing $(\cdot,\cdot)$ in (2.4). 

\proclaim
{Proposition 2.3} If $\bl \in \fg,$ then
$$\eqalign{
&\Pi^{*}_{\frak k} \bl = \bl_{-} - \bl^{T}_{+} \cr
&\Pi^{*}_{\frak l} \bl = \bl_{+} + \bl_{0} + \bl^{T}_{+}.\cr}\eqno(2.11)
$$
\endproclaim

\demo
{Proof}  For all $\bl,\ba\in \fg,$ we have the obvious relation
$(\bl, \ba_{+}) = (\bl_{-}, \ba_{+}) =(\bl_{-},\ba).$  Similarly,
$(\bl,\ba^{T}_{+})=(\bl_{+}, \ba^{T}_{+}) = (\bl^{T}_{+},\ba).$
Hence the formula for $\Pi^{*}_{\frak k}$ follows.  The formula
for $\Pi^{*}_{\frak l}$ is now obtained from
$\Pi^{*}_{\frak k} + \Pi^{*}_{\frak l} =1.$

\pf
\enddemo

We will equip $\fg^{*}_{R}\simeq \fg$ with the Lie-Poisson
structure
$$\{\,F_1,F_2\,\}_{R}(\bl) =(\bl, [dF_1(\bl),dF_2(\bl)]_{R})\eqno(2.12)$$
where $F_1$, $F_2 \in C^{\infty}(\fg^{*}_{R}),$ and
$dF_{i}(\bl)\in \fg$ is defined by the formula
${d\over dt}\big|_{t=0} F_{i}(\bl + t\bl^{\prime}) = (dF_{i}(\bl), \bl^{\prime}),\,
i=1,2.$

The following result is a consequence of standard classical r-matrix theory. 
(See  \c{STS} and \c{RSTS} for the general theory.)

\proclaim
{Proposition 2.4} (a) The Hamiltonian equations of motion generated
by $F \in C^{\infty}(\fg^{*}_{R})$ is given by
$$\dot \bl = {1\over 2}[\,R(dF(\bl)),\bl\,] -{1\over 2}R^{*}[\,\bl, dF(\bl)\,].
\eqno(2.13)$$
In particular, for the Hamiltonian $H_{j}(\bl) = {1\over 2(j+1) }tr(\bl^{j+1})$,
$j= 1,2,\ldots$, the corresponding equation is the Lax equation
$$\dot \bl = {1\over 2} [\,\bl,\,\,\Pi_{\frak k} \,\bl^{j}\,].\eqno(2.14)$$
\smallskip
\noindent (b) The family of functions $H_{j}(\bl)$, $j=1,2,\ldots$ Poisson
commute with respect to $\{\cdot,\cdot\}_{R}$.
\endproclaim
Let
$$\frak p = \lbrace \bl\in \fg\mid \bl = \bl^{T} \rbrace.\eqno(2.15)$$

\proclaim
{Corollary 2.5} (a) $\frak p$ is a Poisson submanifold of
$(\fg^{*}_{R}, \{\cdot,\cdot\}_{R})$. Hence eqn. (2.14) with
$\bl\in \frak p$ is Hamiltonian with respect to the
induced Poisson structure on $\frak p.$
\endproclaim

{\it Usage. \/}\,\, The flows defined by (2.14) will be called collectively 
the $(-)$ Toda flows.
On the other hand, the $(-)$ Toda flow will refer to the $j=1$ case
in (2.14).  In this work, we will also consider
the $(+)$ Toda flow which is defined by the equation
$\dot \bl = {1\over 2} [\,\Pi_{\frak k} \,\bl\,\, , \bl].$

Our next goal in this section is to show how to solve
the $(\pm)$ Toda flow.   In order to do so, we have
to describe the Lie groups which integrates the Hilbert
Lie algebras $\fg$, $\frak l$ and $\frak k.$  
For this purpose, let $GL({\Cal H})$
be the group of invertible operators in $B({\Cal H}),$ 
and let $\bi \in GL({\Cal H})$ denote the identity operator.
We begin by defining
$$ G = GL({\Cal H})\bigcap\, (\bi + \fg).\eqno(2.16)$$ 
Since $\fg$ is a 2-sided ideal in $B({\Cal H})$ (see, for example
\c{RS1}), it is clear that
if $\bi + \ba\in G$, then $(\bi + \ba)^{-1}\in G.$
Thus as in \c{L2}, we can show that $G$ is a Hilbert Lie
group, it is indeed the Hilbert Lie group which integrates
$\fg.$   We will call $G$ the {\sl Hilbert-Schmidt group}.
On the other hand, the Lie subgroup of $G$ which corresponds
to the Lie algebra $\frak k$ is given by
$${\Cal K}= \{\, k\in G\mid kk^{T} = k^{T} k = \bi\, \}.\eqno(2.17)$$
In order to introduce the Lie subgroup of $G$ which integrates $\frak l,$
we need some preparation.   Our next result is a discrete version
of Lemma 1, Section 2.7 of \c{Sm}.  We can prove it inductively
as in \c{Sm}.  For this reason, we will omit its details.

\proclaim
{Lemma 2.6} Let $\ba= (A_{ij})_{i,j=1}^{\infty}\in \frak l$ and 
$u=(u(1), u(2),\cdots)\in {\Cal H}$.  If 
$$u_{n} = \ba^{n} u, \quad n=1,2,\cdots , \eqno(2.18)$$
then
$$|u_{n}(j)|\leq \frac{A_{1}(j)\vert|u\vert|}{[(n-1)!]^{1\over 2}}
  \left(\sum_{k=1}^{j} A_{1}(k)^{2}\right)^{{\frac{1}{2}}(n-1)}\eqno(2.19)$$
for all $j\in \Bbb N,$
where 
$$A_{1}(j) = \left(\sum_{k=1}^{j} a_{jk}^{2} \right)^{\frac{1}{2}},\,\, j\in
\Bbb N.\eqno(2.20)$$
\endproclaim

\proclaim
{Proposition 2.7} The set
$${\Cal L} =\{\,g\in \bi + \frak l\mid g_{ii} > 0\,\, \hbox{for all}\,\,i\,\}
\eqno(2.21)$$
with the induced group operation of $G$ is a Lie subgroup 
of $G$ which integrates $\frak l.$
\endproclaim

\demo
{Proof} It is clear that the group operation of $G$ is closed 
on ${\Cal L}$.  On the other hand, if $\ba\in \frak l$, we can
show that the Neumann series $\sum_{n=0}^{\infty} (-1)^{n}\ba^{n}$
converges.  To see this, take a nonzero 
vector $u=(u(1), u(2),\cdots)\in {\Cal H}.$
Then from the inequality in (2.19), we have
$$\eqalign{
\vert|\ba^{n} u\vert|^{2} & \leq \sum_{j=1}^{\infty}
\frac{A_{1}(j)^{2}\vert|u\vert|^{2}}{(n-1)!}
  \left(\sum_{k=1}^{j} A_{1}(k)^{2}\right)^{n-1}\cr
 & = \frac{\vert|u\vert|^{2}} {n!}\left(\sum_{j=1}^{\infty} A_{1}(j)^{2}\right)^{n}\cr  & = \frac{\vert|u\vert|^{2}} {n!} \vert|\ba\vert|_{2}^{2n}.\cr}\eqno(2.22)$$
from which we obtain the estimate
$$\vert|\ba^n\vert| \leq \frac{\vert|\ba\vert|^{n}_{2}}{\sqrt{n!}}.\eqno(2.23)$$
The convergence of the Neumann series is now clear from (2.23).
Thus $\bi + \ba$ is invertible with 
$(\bi + \ba)^{-1} = \sum_{n=0}^{\infty} (-1)^{n} \ba^{n}\in {\Cal L}.$
This shows that ${\Cal L}$ is a subgroup of $G.$  As ${\Cal L}$
is clearly a submanifold of $G$, this completes the proof of
the assertion.
\pf
\enddemo

Our next result is the global version of the direct sum decomposition
in Proposition 2.2.

\proclaim
{Proposition 2.8} Suppose $\bi + \ba \in G$, then $\bi + \ba$
has a unique factorization
$$\bi + \ba = b_{-}b^{-1}_{+}\eqno(2.24)$$
where $b_{-}\in {\Cal L}$ and $b_{+}\in {\Cal K}.$
\endproclaim

\demo
{Proof} The factorization problem in (2.24) is equivalent to
$(\bi + \ba)^{T} = b_{+}b_{-}^{T},$ where $b_{-}\in {\Cal L}$ and 
$b_{+}\in {\Cal K}.$  Now we can certainly obtain a unique orthogonal
$b_{+}\in GL({\Cal H})$ and a unique lower triangular $b_{-}\in GL({\Cal H})$
with $(b_{-})_{ii} >0$ by applying the Gram-Schmidt orthogonalization 
process to the vectors $(\bi + \ba)^{T} e_1, (\bi + \ba)^{T} e_2, \cdots.$   
To complete the proof, it suffices to show that $b_{-}\in {\Cal L}.$
To this end, note that $(\bi + \ba)(\bi + \ba)^{T} = b_{-}b_{-}^{T}.$
Since $\fg$ is a 2-sided ideal in $B({\Cal H})$ which is closed
under the operation of taking the transpose, we can rewrite
the above relation in the form
$$\bk = b_{-} - (b_{-}^{-1})^{T}\eqno(2.25)$$
where
$$\bk = ((\bi + \ba)(\bi + \ba)^{T} -\bi)(b_{-}^{-1})^{T}\in \fg.\eqno(2.26)$$
Now, from $\vert|\bk\vert|^{2}_{2} <\infty$ and the relation in (2.25),
we infer that 
$$\sum_{j<i} (b_{-})^{2}_{ij} + \sum_{i<j} (b^{-1}_{-})^{2}_{ji}
  + \sum_{i=0}^{\infty} \frac{((b_{-})^{2}_{ii} -1)^{2}}{(b_{-})^{2}_{ii}} <\infty.
\eqno(2.27)$$
But as
$$\eqalign{
&\frac{((b_{-})^{2}_{ii} -1)^{2}}{(b_{-})^{2}_{ii}}-((b_{-})_{ii}-1)^{2}\cr
=\,\, &\frac{((b_{-})_{ii}-1)^{2}(2(b_{-})_{ii}^{2}+1)}{(b_{-})^{2}_{ii}}\cr
\geq\,\, &0,\cr}\eqno(2.28)$$
it follows on using (2.28) in (2.27) that
$$\vert|b_{-}-\bi\vert|^{2}_{2} = \sum_{j<i} (b_{-})^{2}_{ij} +
  \sum_{i=1}^{\infty} ((b_{-})_{ii}-1)^{2} <\infty,\eqno(2.29)$$
as desired.

\pf
\enddemo

We are now ready to give the solution to the $(\pm)$ Toda flow.
As the proof is quite standard, we refer the reader to 
Theorem 3.2 and Remark 3.3 (b) in \c{L2}.
(See \c{RSTS} for the general theory of the factorization method.)

\proclaim
{Theorem 2.9} Let $\bl_{0}\in \fg,$ and let $b_{-}(t)\in {\Cal L}$,
$b_{+}(t)\in {\Cal K}$ be the unique solution of the factorization
problem
$$exp\left(\pm{1\over 2}t \bl_{0}\right) = b_{-}(t)b_{+}(t)^{-1}.\eqno(2.30)$$
Then for all $t,$
$$\bl(t) = b_{+}^{-1}(t)\bl_{0}b_{+}(t)= b_{-}^{-1}(t)\bl_{0}b_{-}(t)\eqno(2.31)$$
solves the initial value problem
$$\dot \bl = \pm {1\over 2}[\Pi_{\frak k} \bl, \bl], \quad 
  \bl(0) = \bl_{0}.\eqno(2.32)$$
\endproclaim

We next give the first result on the long time
behaviour of the $(\pm)$ Toda flow when the initial
data $\bl_{0}\in \frak p.$  It is in fact just a special case 
of Proposition 5 in Section 2 of \c{DLT1}. (The proof
is a modification of Moser's argument in \c{Mo}.)

\proclaim
{Proposition 2.10} Let $\bl(t)$ be the solution of 
$\dot \bl = \pm {1\over 2} [\,\Pi_{\frak k} \,\bl, \bl\,\,],$
\,\, $\bl(0)=\bl_{0}\in \frak p.$  Then $\bl(t)$ converges
strongly  to a diagonal operator 
$\bl^{\pm}(\infty)=diag(\alpha^{\pm}_1, \alpha^{\pm}_2, \cdots)$
with $\alpha^{\pm}_i$ belonging to the spectrum $\sigma(\bl_{0})$ 
of $\bl_{0}$ as $t\to \infty.$
\endproclaim

\noindent{\bf Remark 2.11} Recall that $\bl(t)$ converges
{\it strongly \/} to $\bl^{\pm}(\infty)$ as $t\to \infty$ means 
$\vert|\bl(t)u -\bl^{\pm}(\infty)u\vert|\to 0$ for each
$u\in {\Cal H}$ (see \c{RS1}).  Since ${\Cal H}$ is infinite dimensional,
this notion of convergence is weaker than norm convergence, 
so in general the spectrum of $\bl^{\pm}(\infty)$ can shrink.
(See VIII.7 of \c{RS1} for a discussion of such matters.)
As the reader will see, this is indeed what happens 
in Section 4 below.

\medskip

In the rest of the section, we will describe the symplectic
leaves of the Lie-Poisson structure $\{\cdot,\cdot \}_{R}$ which
are given by the coadjoint orbits of the infinite dimensional
Lie group $G_{R}$  which integrates $\fg_R.$ In particular,
we will consider the coadjoint action of $G_R$ on the class $\frak p_{*}$ of
{\it semiseparable \/} operators $\bl\in \fg.$   By definition, 
a Hilbert-Schmidt
operator $L = (L_{ij})_{i,j=1}^{\infty}\in \frak p_{*}$ if and only if
$$L_{ij} =\cases u_iv_j, & i\leq j \\
                u_jv_i, & i>j, \endcases\eqno(2.33)$$
where $u=(u_1,u_2,\cdots)$, $v=(v_1,v_2,\cdots)$ are sequences
of real numbers, which are not necessarily in $l^{+}_{2}.$
The Lie group $G_{R}$ can be described in the following
way (cf. \c{DLT2}): the underlying manifold is $G$, but
now the group operation is defined by
$$g\ast h \equiv g_{-}hg_{+}^{-1}\eqno(2.34)$$
where $g = g_{-}g_{+}^{-1}$ is the unique factorization into
$g_{-}\in {\Cal L}$ and $g_{+}\in {\Cal K}.$  Moreover,
the coadjoint action of $G_{R}$ on  $\fg^{*}_{R}\simeq \fg$
is given by
$$Ad^{*}_{G_{R}}(g^{-1})\bl = \Pi^{*}_{\frak l}(g_{-} \bl g^{-1}_{-})
  + \Pi^{*}_{\frak k}(g_{+}\bl g^{-1}_{+})\eqno(2.35)$$
and the orbits of this action are the symplectic leaves
of $\{\cdot,\cdot \}_{R}.$

\proclaim
{Proposition 2.12} The class $\frak p_{*}\subset \frak p$ of
Hilbert-Schmidt operators in ${\Cal H}$ which are semiseparable
is invariant under $Ad^{*}_{G_R}.$
\endproclaim

\demo
{Proof} From (2.35), we have
$Ad^{*}_{G_{R}}(g)\bl = \Pi^{*}_{\frak l}(g^{-1}_{-}\bl g_{-})$
for $\bl\in \frak p_{*}.$
If
$$L_{ij} =\cases u_i v_j, & i\leq j \\
                u_j v_i, & i>j, \endcases$$
for some sequences of real numbers
$u=(u_1,u_2,\cdots)$ and $v=(v_1,v_2,\cdots),$
a straightforward computation shows that
$$(g^{-1}_{-}\bl g_{-})_{ij} = (g^{-1}_{-}u)_{i}(g^{T}v)_{j},\quad i\leq j$$
where we have used the fact that $g_{-}$ is lower triangular.
Therefore the assertion follows from the formula for $\Pi^{*}_{\frak l}$
in (2.11)

\pf
\enddemo

From this result, it follows that if the initial data $\bl_{0}$ of
(2.32) is in $\frak p_{*}$, then $\bl(t)\in \frak p_{*}$ for
all $t.$   In the next section, the reader will see that we will be
dealing with the $(\pm)$ Toda flow on some rather special
semiseparable operators which are related to the CH equation.

\bigskip
\bigskip

\subhead
3. \ A class of low-regularity solutions of the Camassa-Holm equation
\endsubhead

\bigskip

In this section, we will consider a class of weak solutions of the
CH equation
$$u_t -u_{xxt} + 3uu_x = 2u_{x}u_{xx} + uu_{xxx},\eqno(3.1)$$
of the form
$$u(x,t) = \frac{1}{2} \sum_{j=1}^{\infty} e^{-|x-q_{j}(t)|} p_{j}(t),\eqno(3.2)$$
where $p_j(t) \neq 0$ for all $j\in \Bbb N$ and such that 
$p_{j}(t)\to 0$ sufficiently fast as $j\to 0.$   To be more
precise, we assume (for small values of $t$) that
$q(t) = (q_1(t),q_2(t),\cdots)\in l^{+}_{\infty},$
while  $p(t) = (p_1(t),p_2(t),\cdots)\in l^{+}_{1,2}.$   
Here $l^{+}_{\infty}$ and $l^{+}_{1,2}$ are Banach spaces defined as follows:
$$\eqalign{&
l^{+}_{\infty} =\lbrace q = (q_1, q_2, \cdots)\mid \vert|q\vert|_{\infty} = 
sup_{j}\, |q_j| <\infty\rbrace, \cr 
&l^{+}_{1,2} =\lbrace p = (p_1,p_2,\cdots)\mid
\vert|p\vert|_{1,2} = \sum_{j=1}^{\infty} j^{2} |p_{j}|<\infty\rbrace.\cr}\eqno(3.3)$$
Following \c{BSS}, we rewrite the CH equation (3.1) in the form
$$m_{t} + (mu)_{x} + mu_{x} =0, \quad m = u - u_{xx}.\eqno(3.4)$$
Then the solution in (3.2) corresponds to the measure
$$m(x,t) = \sum_{j=1}^{\infty} e^{-|x-q_{j}(t)|}p_{j}(t)\, \delta(x-q_{j}(t)).\eqno(3.5)$$
Therefore, if we mimic the calculation in \c{BSS}, we find that
$u(x,t)$ and $m(x,t)$ satisfy the equation in (3.4) in a weak sense 
if and only
if
$$\eqalign{
&\dot{q}_{j} = \frac{1}{2} \sum_{k=1}^{\infty} e^{-|q_j -q_k|} p_k,\cr
&\dot{p}_{j} = \frac{1}{2} p_j \sum_{k=1}^{\infty} sgn(q_j -q_k)
   e^{-|q_j -q_k|} p_k, \quad j\in \Bbb N\cr}\eqno(3.6)$$
where we adopt the convention that $sgn\, 0 = 0.$
Note that our assumptions above means that we are considering these
equations in the Banach space direct sum $l^{+}_{\infty}\oplus l^{+}_{1,2},$
equipped with the norm 
$\vert|(q,p)\vert| = \vert|q\vert|_{\infty} + \vert|p\vert|_{1,2}.$
Clearly, the signs of the $p_j$'s are preserved as long as no
blowup occurs.   In this work, we will focus on the case where
$p_j >0$ for all $j.$   Indeed, we will restrict our attention
to the two sectors
$${\Cal S_{-}} = \lbrace (q,p)\in l^{+}_{\infty}\oplus l^{+}_{1,2} \mid
   q_1 < q_2 < \cdots, p_j > 0 \,\, \hbox{for all}\,\, j \rbrace,\eqno(3.7)$$
and 
$${\Cal S_{+}} = \lbrace (q,p)\in l^{+}_{\infty}\oplus l^{+}_{1,2} \mid
   q_1 > q_2 > \cdots, p_j > 0 \,\, \hbox{for all}\,\, j \rbrace\eqno(3.8)$$
for the most part.  At the end of the paper, we will show how to
adapt our analysis to other sectors which are defined by a restricted
class of permutations of $\Bbb N.$

\proclaim 
{Proposition 3.1} Suppose $(q^0,p^0)\in {\Cal S_{\pm}}$.
Then the initial value problem
$$\eqalign{
&\dot{q}_{j} = \frac{1}{2} \sum_{k=1}^{\infty} e^{-|q_j -q_k|} p_k,\cr
&\dot{p}_{j} = \frac{1}{2} p_j \sum_{k=1}^{\infty} sgn(q_j -q_k)
   e^{-|q_j -q_k|} p_k\cr
&\phantom{ab} = \pm \frac{1}{2} p_j \sum_{k=1}^{\infty} sgn(k-j)
   e^{-|q_j -q_k|} p_k,\cr
&q_j(0) = q^{0}_j,\,\,p_j(0) = p^{0}_j, \,\,\,\, j\in \Bbb N\cr}\eqno(3.9)$$ 
has a unique global solution in ${\Cal S_{\pm}}.$
\endproclaim

\demo
{Proof} For $(q,p)\in {\Cal S_{\pm}}$, put
$$f(q,p) = (f_1(q,p), f_2(q,p)) \eqno(3.10)$$
where
$$\eqalign{
&f_1(q,p)= \left(\frac{1}{2} \sum_{j=1}^{\infty} e^{-|q_j -q_k|} p_k \right)_{j=1}^{\infty}, \cr
&f_2(q,p)=\left(\frac{1}{2} p_j \sum_{j=1}^{\infty} sgn(q_j -q_k)
   e^{-|q_j -q_k|} p_k\right)_{j=1}^{\infty} .\cr}\eqno(3.11)$$
We first show $f:{\Cal S_{\pm}}\longrightarrow l^{+}_{\infty}\oplus l^{+}_{1,2}.$
For this purpose (and for later usage), we put
$$\vert|p\vert|_{1} = \sum_{k=1}^{\infty} p_k,\,\,\,\, p= (p_k)_{k=1}^{\infty}\in l^{+}_{1,2}, \,\,p_k>0.\eqno(3.12)$$ 
Then from the expressions for $f_1$ and $f_2$ above, we find that
$$\vert|f(q,p)\vert| \leq \frac{1}{2} \vert|p\vert|_{1}(1 + \vert|p\vert|_{1,2}),\eqno(3.13)$$
as desired.   We next show $f$ is locally Lipschitz.  To do this,
take $(q,p), (\widetilde q, \widetilde p)$ in an open ball centered
at $(q^0,p^0)$ which is contained in ${\Cal S_{\pm}}.$  Then by making
use of the inequality $|e^{-\xi} - e^{-\eta}|\leq |\xi -\eta|$ for
$\xi,\eta >0$ and the triangle inequalities, we have
$$\eqalign{
&\Big|\sum_{k=1}^{\infty} e^{-|q_j -q_k|} p_k -\sum_{k=1}^{\infty} e^{-|\widetilde{q}_j -\widetilde{q}_k|}\widetilde{p}_k\Big|\cr
\leq\, & \sum_{k=1}^{\infty} e^{-|q_j -q_k|}|p_k -\widetilde p_{k}| +
\sum_{k=1}^{\infty} |\widetilde p_k||e^{-|q_j -q_k|}-e^{-|\widetilde{q}_j -\widetilde{q}_k|}|\cr
\leq\, & \sum_{k=1}^{\infty} |p_k -\widetilde p_k| +  \sum_{k=1}^{\infty}|\widetilde 
p_k|(|q_j-\widetilde q_j| + |q_k -\widetilde q_k|)\cr
\leq\, & \sum_{k=1}^{\infty} |p_k -\widetilde p_k| + \vert|\widetilde{p}\vert|_{1}|q_j -\widetilde q_j| + \vert|\widetilde{p}\vert|_{1}\,\vert|q-\widetilde{q}\vert|_{\infty}.\cr}\eqno(3.14)$$
Consequently,
$$\eqalign{
&\vert|f_1(q,p) - f_1(\widetilde{q}, \widetilde{p})\vert|_{\infty}\cr
\leq\, &\frac{1}{2} \vert|p -\widetilde{p}\vert|_{1,2} + \vert|\widetilde{p}\vert|_{1}\vert|q-\widetilde{q}\vert|_{\infty}\cr}\eqno(3.15)$$
On the other hand, by using the fact that $sgn(q_j -q_k) = sgn(\widetilde{q}_j-
\widetilde{q}_k) = \pm sgn(k-j)$ and the estimate in (3.14), we find
$$\eqalign{
&\vert|f_2(q,p) - f_2(\widetilde{q}, \widetilde{p})\vert|_{1,2}\cr
\leq\, & \frac{1}{2} \sum_{j=1}^{\infty} j^{2} \sum_{k=1}^{\infty} |e^{-|q_j -q_k|} p_k -
 e^{-|\widetilde{q}_j -\widetilde{q}_k|}\widetilde{p}_k|\,p_j\cr
& +  \frac{1}{2} \sum_{j=1}^{\infty} j^{2} \sum_{k=1}^{\infty} e^{-|\widetilde{q}_j -\widetilde{q}_k|} \,\widetilde{p}_{k}|p_j-\widetilde{p}_j|\cr
\leq\, & \vert|p\vert|_{1,2}\left(\frac{1}{2}\vert|p-\widetilde{p}\vert|_{1,2} + \vert|\widetilde{p}\vert|_{1}\vert|q-\widetilde{q}\vert|_{\infty}\right) + \frac{1}{2}\vert|\widetilde{p}\vert|_{1}\vert|p-\widetilde{p}\vert|_{1,2}.\cr}\eqno(3.16)$$ 
Therefore, upon combining (3.15) and (3.16), we obtain
$$\eqalign{
& \vert|f(q,p) - f(\widetilde{q},\widetilde{p})\vert|\cr
\leq &C(\vert|p\vert|_{1,2}, \vert|\widetilde{p}\vert|_{1})\,\vert|(q,p)-(\widetilde{q},\widetilde{p})\vert|.\cr}\eqno(3.17)$$
Finally, to establish global existence, let us suppose the solution
$(q(t), p(t))$ exists for $0\leq t\leq T$ for some $T >0$.   We will
establish an a priori estimate for $\vert|(q(t),p(t))\vert|$.  To do
so, observe that $P = \sum_{j=1}^{\infty} p_{j}(t)$ is a conserved
quantity.  Hence the equations for $q_j(t)$ gives 
$|\dot q_j(t)| \leq \frac{1}{2} P$ from which we obtain
the estimate 
$$\vert|q(t)\vert|_{\infty} \leq \vert|q(0)\vert|_{\infty} + {1\over 2}Pt.\eqno(3.18)$$
Similarly, the equations for $p_j(t)$ gives 
$|\dot p_j(t)| \leq {1\over 2} p_j(t)P$ from which
we find $p_j(t) \leq p_j(0)\, e^{{1\over 2} Pt}.$ Therefore,
$$\vert|p(t)\vert|_{1,2} \leq \vert|p(0)\vert|_{1,2}\, e^{{1\over 2}Pt}.\eqno(3.19)$$
Consequently, on combining (3.18) and (3.19), we conclude that
$$\vert|(q(t),p(t))\vert| \leq \vert|q(0)\vert|_{\infty} + {1\over 2} Pt
  + \vert|p(0)\vert|_{1,2}\, e^{{1\over 2}Pt}\eqno(3.20)$$
for $0\leq t\leq T$.  This completes the proof.
\pf
\enddemo

Our next result relates the equations in (3.6) with $(q,p)\in {\Cal S}_{\pm}$
to the $(\pm)$
Toda flow and gives the spectral properties of the Lax operator.
  
\proclaim
{Proposition 3.2} For $(q,p)\in {\Cal S_{\pm}}$, define the operator
$\bl(q,p)=(L_{ij}(q,p))_{i,j=1}^{\infty}$ on $l^{+}_{2}$ by
$$L_{ij}(q,p) = {1\over 2} e^{-{1\over 2}|q_i-q_j|}\sqrt{p_ip_j},\,\,i,j\geq 1.
\eqno(3.21)$$
Then
\smallskip
\noindent (a) $\bl(q,p)$ is a positive, semiseparable trace-class operator.
Indeed, 
$$L_{ij}(q,p) =\cases u_i(q,p)v_j(q,p), & i\leq j \\
                u_j(q,p)v_i(q,p), & i>j, \endcases\eqno(3.22 a)$$
for $(q,p)\in {\Cal S_{-}},$ while
$$L_{ij}(q,p) =\cases u_j(q,p)v_i(q,p), & i\leq j \\
                u_i(q,p)v_j(q,p), & i>j, \endcases\eqno(3.22 b)$$
for $(q,p)\in {\Cal S}_{+},$ where
$$u_i(q,p) = \frac{1}{\sqrt{2}} e^{{1\over 2}q_i}\sqrt{p_i},\,\,\,
  v_i(q,p) = \frac{1}{\sqrt{2}} e^{-{1\over 2}q_i}\sqrt{p_i}, \,\,\,i\geq 1.
\eqno(3.23)$$
\smallskip
\noindent (b) If $f = (f(1),f(2),\cdots )$ is an eigenvector of $\bl(q,p)$, 
then $f(1)\neq 0.$
\smallskip
\noindent (c) The eigenvalues of $\bl(q,p)$ are simple and 
ker\,$(\bl(q,p)) =\{0\}$.
\smallskip
\noindent (d) If $(q,p)$ evolves under (3.6), then 
$$\dot \bl(q,p) = \cases
{1\over 2} [\,\bl(q,p), \Pi_{\frak k} \bl(q,p)\,],& (q,p)\in {\Cal S_{-}}\\
{1\over 2} [\Pi_{\frak k} \bl(q,p)\, \bl(q,p) \,], & (q,p)\in {\Cal S_{+}}.
\endcases\eqno(3.24)$$
\endproclaim

\demo
{Proof} We will give the proof for $(q,p)\in {\Cal S}_{-},$
the arguments for the other case are similar.
\newline
(a) It is clear from the definition of $\bl(q,p)$ that
$L_{ij}(q,p)$ is of the form given in (3.22a) with $u_i(q,p), v_i(q,p)$ 
defined in (3.23).  Moreover, it follows from the natural ordering
of the $q_i$'s that
$$0 < \frac{u_1(q,p)}{v_1(q,p)} < \frac{u_2(q,p)}{v_2(q,p)}<\cdots.\eqno(3.25)
$$
To show that $\bl(q,p)$ is positive, let us denote by $\bl^{(n)}(q,p)$
the $n\times n$ matrix in which $L^{(n)}_{ij}(q,p) = L_{ij}(q,p),\,\, i,j=1
\cdots, n.$  Clearly, $\det(\bl^{(1)}(q,p)) = p_1 >0.$   For $n >1$,
it follows from \c{GK} [eqn.(29) on p.78] that
$$\eqalign{
\det(\bl^{(n)}(q,p))&= u_1(q,p) v_n(q,p)\prod_{j=1}^{n-1} \det \pmatrix
 v_i(q,p) & v_{i+1}(q,p)\cr
 u_i(q,p) & u_{i+1}(q,p)\cr
 \endpmatrix \cr
&= v^{2}_{1}(q,p)\cdots v^{2}_{n}(q,p)\prod_{j=1}^{n} \left(\frac{u_j(q,p)}{v_{j}(q,p)}
   -\frac{u_{j-1}(q,p)}{v_{j-1}(q,p)}\right)\cr}\eqno(3.26)$$
where we formally set $\frac{u_{0}(q,p)}{v_{0}(q,p)}=0.$  Hence we
conclude from (3.25) that $\det(\bl^{(n)}(q,p))>0.$  Consequently,
$\bl(q,p)$ is positive.   Finally, the fact that $\bl(q,p)$ is
trace-class follows as we have
$\sum_{j=1}^{\infty} (e_j, \bl(q,p)e_j )=\sum_{j=1}^{\infty} p_j <\infty.$
\newline
(b) Suppose $\bl(q,p)f = \lambda f$. Writing this out in terms of
components, we have
$$\eqalign{
&u_1(q,p)(v_1(q,p)f(1)+v_2(q,p)f(2)+v_3(q,p)f(3)+\cdots) =\lambda f(1),\cr
&u_1(q,p)v_2(q,p)f(1)+u_2(q,p)(v_2(q,p)f(2)+v_3(q,p)f(3)+\cdots) =\lambda f(2),\cr
&u_1(q,p)v_3(q,p)f(1)+u_2(q,p)v_3(q,p)f(2)+u_3(q,p)(v_3(q,p)f(3)+\cdots) =\lambda
f(3),\cr
&\vdots\cr}\eqno(3.27)$$
If $f(1) =0,$ then it follows from the first equation of (3.27) that
$$v_2(q,p)f(2)+v_3(q,p)f(3)+\cdots =0.$$
Substitute this into the second equation of (3.27), we find $f(2)=0$.
Hence $v_3(q,p)f(3)+\cdots =0.$   When we substitute this into
the third equation of (3.27), we obtain $f(3)=0.$  Proceeding
inductively, it is easy to see that $f=0$, a contradiction
to the assumption that $f$ is an eigenvector. 
\newline
(c) Suppose there exist independent eigenvectors $f$ and $g$ of
$\bl(q,p)$ corresponding to the eigenvalue $\lambda.$  Since
$f(1)$, $g(1)\neq 0$, we can find $c_1, c_2\in \Bbb R\setminus \{0\}$
such that $c_1 f(1) + c_2 g(1) =0.$  But on the other hand,
$c_1 f + c_2 g$ is obviously an eigenvector of $\bl(q,p)$,
which is impossible as $c_1 f(1) + c_2 g(1) =0.$  To show that
ker\,$(\bl(q,p)) =\{0\},$ suppose 
$\bl(q,p)h = 0$. Writing this out in terms of
components, we get
$$\eqalign{
&u_1(q,p)(v_1(q,p)h(1)+v_2(q,p)h(2)+v_3(q,p)h(3)+\cdots) = 0,\cr
&u_1(q,p)v_2(q,p)h(1)+u_2(q,p)(v_2(q,p)h(2)+v_3(q,p)h(3)+\cdots) = 0,\cr
&u_1(q,p)v_3(q,p)h(1)+u_2(q,p)v_3(q,p)h(2)+u_3(q,p)(v_3(q,p)h(3)+\cdots) = 0,\cr
&\vdots\cr}\eqno(3.28)$$
Next, we multiply the second equation of (3.28) by $-\frac{u_1(q,p)}{u_2(q,p)},$
and add it to the first equation, this gives 
$$v_1(q,p)v_2(q,p)\left(\frac{u_2(q,p)}{v_2(q,p)}-\frac{u_1(q,p)}{v_1(q,p)}
\right)\frac{u_1(q,p)}{v_1(q,p)}h(1)=0.\eqno(3.29)$$
Thus it follows from (3.29) and (3.25) that $h(1) =0.$  Clearly, if
we proceed inductively and make use of (3.25), the conclusion is
that $h=0,$ as required.
\newline
(d) See Remark 3.3 (a) below.
\pf
\enddemo

\noindent{\bf Remark 3.3} (a) If we replace $\infty$ by a positive
integer $N$ in (3.2), we obtain the multipeakon
solutions of the CH equation \c{CH}, \c{BSS} if $p_j(t) >0$
for $j=1,\cdots, N.$  In that case,
the Lax pair of the Hamiltonian equations for $(q,p)\in \Bbb R^{2N}$
with $p_j >0$ for all $j$ was discovered in \c{CF} in a remarkable
calculation.   On the other hand,
in the sector where the $q_j$'s satisfy the natural ordering
$q_1 < q_2<\cdots<q_N$, the realization
that the Lax equation is just a special case of the Toda flows
on $N\times N$ matrices was pointed out in \c{RB}. In this regard, the 
calculation which leads to (3.24) is just an extension of the one in
\c{CF}.  We should mention, however, that the r-matrix in
\c{RB} is not the appropriate one to use from the point of view
of Poisson geometry.  Indeed, it is easy to check that the
set of $N\times N$ symmetric matrices is not even a Poisson
submanifold of the Lie-Poisson structure associated with the 
corresponding $R$-bracket. 
\newline
(b) In the case of the peakons lattice in \c{RB}, the Lax
operator is invertible.(This also follows from (3.26) above.)
In our case, although $0$ is not an eigenvalue
of $\bl(q,p)$ by Proposition 3.2 (c) above, however, it is in the 
essential spectrum $\sigma_{ess}(\bl(q,p))$ (indeed, 
$\sigma_{ess}(\bl(q,p)) =\{0\}).$ Hence $\bl(q,p)$ is not
invertible.
\medskip

We close this section with the following result which is a consequence of 
Proposition 2.10, Proposition 3.2 and the equation for $q_j$ in
(3.6).

\proclaim
{Proposition 3.4} Let $(q(t),p(t))$ be a solution of (3.6) 
in the sector ${\Cal S_{\pm}}$. Then 
\smallskip
\noindent (a) $\bl(q(t),p(t)) \to diag(\alpha^{\pm}_{1},\alpha^{\pm}_{2},\cdots)$ 
strongly as $t\to\infty$ where 
$\alpha^{\pm}_{i} \in \sigma(\bl(q(0),p(0))$ and hence 
 $\lim_{t\to \infty} p_j(t) = 2 \alpha^{\pm}_j$ for each $j\in \Bbb N,$
\newline
(b) $\dot q_{j}(t) >0$ for each $j\in \Bbb N$ and 
$\lim_{t\to \infty} e^{-\frac{1}{2}|q_{j}(t) -q_{k}(t)|}\sqrt{p_{j}(t)p_{k}(t)}=0$
whenever $j\neq k.$
\endproclaim

\noindent{\bf Remark 3.5} (a) From Proposition 3.4 (b) above, 
we conclude that the peakons are traveling to the right.
\newline
\noindent (b) In Sections 4 and 5, we will show that the peakons
separate out, i.e., the $q_j(t)$'s have the scattering behaviour.
However, in contrast to the semi-infinite Toda lattice \c{L1}, 
this does not follow 
immediately from the long time behaviour of $\bl(q(t), p(t)).$   
This is clear from Proposition 3.4 (b) above as
$\lim_{t\to \infty}  e^{-\frac{1}{2}|q_{j}(t) -q_{k}(t)|} = 0,\,\,j\neq k$
does not follow automatically from
$\lim_{t\to \infty} e^{-\frac{1}{2}|q_{j}(t) -q_{k}(t)|}\sqrt{p_{j}(t)p_{k}(t)}=0.$
Indeed, for $(q(t), p(t))\in {\Cal S}_{-}$, we will show that
$\lim_{t\to\infty} p_j(t) =0$ for all $j\in \Bbb N$, so even 
the explicit values of the $\alpha^{-}_j$'s are of
no help in this case in establishing the scattering behaviour.

\bigskip
\bigskip

\subhead
4. \ Long time behaviour in the sector ${\Cal S}_{-}$
\endsubhead
\bigskip

Let $(q(t), p(t))$ be the solution of (3.6) with 
$(q(0), p(0))= (q^0,p^0)\in {\Cal S}_{-}$ and let
$\sigma (\bl(q^0,p^0))\setminus \{0\} =\{\lambda_i\}_{i=1}^{\infty}.$
In view of Proposition 3.2, we will order the eigenvalues as follows:
$$0 < \cdots < \lambda_3 < \lambda_2 < \lambda_1.\eqno(4.1)$$
We will also take the normalized eigenvectors 
$\phi_{1}(t), \phi_{2}(t),\cdots$ of $\bl(q(t),p(t))$ to be such that 
$\phi_{k}(1,t) >0,\,\,k=1,2, \cdots.$

Now it follows from the same proposition that $\bl(q(t),p(t))$
is one-to-one.  Hence $\bl(q(t), p(t))$ has a left inverse
$\bj(q(t),p(t))$ which is an unbounded operator defined on
the dense linear subspace Ran\,$\bl(q(t),p(t))$ of ${\Cal H}.$
In order to give the formula
for  $\bj(q(t),p(t)),$ set
$$e_{j}(t) = e^{-{\frac{1}{2}(q_{j+1}(t)-q_{j}(t))}},\,\, j\in \Bbb N.\eqno(4.2)$$ 

\proclaim
{Proposition 4.1} The matrix of $\bj(q(t),p(t))$ is tridiagonal:
$$\bj(q(t),p(t)) = \pmatrix
a_1(t) & -b_1(t) & 0 & \cdots\cr
-b_1(t) & a_2(t) & -b_2(t) & \ddots \cr
0 & -b_2(t) & a_3(t) & \ddots\cr
\vdots & \ddots & \ddots & \ddots \cr
\endpmatrix \eqno(4.3)$$
with
$$\eqalign{
&a_{j}(t) = \frac{2}{p_j(t)}\frac{1-e^{2}_{j-1}(t)e^{2}_{j}(t)}{(1-e^{2}_{j-1}(t))
  (1-e^{2}_{j}(t))},\cr
&b_{j}(t) = \frac{2}{\sqrt{p_{j}(t)p_{j+1}(t)}}\frac{e_{j}(t)}{1-e^{2}_{j}(t)},\,\,
  j\in \Bbb N\cr}\eqno(4.4)$$
where we formally set $e_{0}(t)=0.$

Moreover,
$$\eqalign{ e_{j}(t)\frac{\phi_{k}(j+1,t)}{\sqrt{p_{j+1}(t)}}
= & -\frac{e_{j-1}(t)(1-e^{2}_{j}(t))}{1-e^{2}_{j-1}(t)}\frac{\phi_{k}(j-1,t)}
{\sqrt{p_{j-1}(t)}}\cr
 & +
 {1\over 2}(1-e^{2}_{j}(t)) p_{j}(t) \left(a_{j}(t)-\frac{1}{\lambda_{k}}\right)
\frac{\phi_{k}(j,t)}{\sqrt{p_{j}(t)}}\cr}\eqno(4.5)$$
for all $j,k\in \Bbb N.$
\endproclaim

\demo
{Proof} To obtain the formula for $\bj(q(t),p(t))$, we
solve $\bl(q(t),p(t))f = g$ recursively for $f(1), f(2),\cdots$
in terms of the components of $g$.  On the other hand,
from $\bl(q(t),p(t))\phi_{k}(t) = \lambda_{k}\phi_{k}(t),$
we find $\bj(q(t),p(t))\phi_{k}(t) = {1\over \lambda_{k}}\phi_{k}(t).$
Since $\bj(q(t),p(t)$ is given by (4.3), we obtain the recurrence relation
$$ b_{j}(t)\phi_{k}(j+1,t) = -b_{j-1}(t) \phi_{k}(j-1,t) +
 \left(a_{j}(t)-\frac{1}{\lambda_{k}}\right)\phi_{k}(j,t).\eqno(4.6)$$
Therefore, the formula in (4.5) follows upon multiplying both sides of
(4.6) by
${1\over 2}\sqrt{p_{j}(t)}(1-e^{2}_{j}(t))$ and
making use of the formulas in (4.4).
\pf
\enddemo

Our next result shows $\phi_{k}(1,t)$ can be solved explicitly. 

\proclaim
{Lemma 4.2} For each $k\in \Bbb N$, 
$$\phi_{k}(1,t) = \frac{e^{-{1\over 2}\lambda_{k}t} \phi_{k}(1,0)}
  {\left(\sum_{j=1}^{\infty} e^{-\lambda_{j}t} \phi^{2}_{j}(1,0)\right)^{1\over 2}}.
\eqno(4.7)
$$
\endproclaim

\demo
{Proof} Let $\bl(t)= \bl(q(t),p(t))$. Then $\bl(t)$ evolves under the
$(-)$ Toda flow. By Theorem 2.9, for any
$j\in \Bbb N,$
$$\eqalign{
(e_1, \phi_{j}(t)) = &\,\, (e_1, b_{+}(t)^{-1}\phi_{j}(0))\cr
                  = &\,\, ((b_{-}(t)^{-1})^{T} e_1, e^{-{1\over 2}t\bl(0)}\phi_{j}(0))\cr
                  = &\,\, \frac{e^{-{1\over 2}\lambda_{j}t}(e_1, \phi_{j}(0))}
                      {(b_{-}(t))_{11}}.\cr}\eqno(4.8)$$
The assertion therefore follows from (4.8) and the relation
$\sum_{j=1}^{\infty} \phi_{j}^{2}(1,t) =1.$
\pf
\enddemo

As the $\lambda_j$'s accumulate at $0$, it is a difficult problem
to get the asymptotics of $\phi_k(1,t)$ as $t\to\infty$ from
(4.7) above. In the following, we will bypass this difficulty.

\proclaim
{Theorem 4.3} Let $(q(t), p(t))$ be the solution of (3.6) with
$(q(0), p(0))= (q^0,p^0)\in {\Cal S}_{-}$, then
\newline
\noindent (a)  $\lim_{t\to \infty} p_j(t) = 0$ for all $j\in \Bbb N,$
\newline
\noindent (b) $\lim_{t\to \infty} |q_{j}(t) -q_{k}(t)| =\infty$ for all
$j\neq k,$
\newline
\noindent (c) $q_j(t)\to \infty$ as $t\to\infty$ for all $j\in \Bbb N,$
\newline
\noindent (d) $q_j(t) = o(t)$ as $t\to\infty$ for all $j\in \Bbb N.$
\endproclaim

\demo
{Proof} From (4.7), we have
$$\frac{\phi_{r}(1,t)}{\phi_{s}(1,t)} =e^{-{1\over 2}(\lambda_r -\lambda_s)t}
  \frac{\phi_{r}(1,0)}{\phi_{s}(1,0)}\eqno(4.9)$$
for all $r,s\in \Bbb N.$  Therefore,
$$\eqalign{
\frac{\phi_{k}^{2}(1,t)}{p_{1}(t)}
  =&\,\frac{\phi_{k}^{2}(1,t)}{2\sum_{j=1}^{\infty} \lambda_{j}\phi^{2}_{j}(1,t)}\cr
  <&\,\frac{1}{2\lambda_{k+1}} \frac{\phi_{k}^{2}(1,t)}{\phi_{k+1}^{2}(1,t)}\cr
  =&\, c_{k1}\,e^{-(\lambda_{k}-\lambda_{k+1})\,t}\cr}\eqno(4.10)$$
for each $k\geq 1$ where $c_{k1}$ depends only $\bl(q^0,p^0).$  Note that as 
$\lim_{t\to\infty} p_{1}(t) = 2 \alpha^{-}_1 <\infty,$
the above inequality not only gives 
$$\lim_{t\to\infty} \frac{\phi_{k}(1,t)}{\sqrt{p_{1}(t)}}=0,\eqno(4.11)$$ 
but also
$$\lim_{t\to \infty} \phi_{k}(1,t) =0.\eqno(4.12)$$
Consequently, we conclude that
$$\lim_{t\to\infty} p_{1}(t) = 2\lim_{t\to\infty}\sum_{j=1}^{\infty} \lambda_{j} \phi^{2}_{j}(1,t)=0\eqno(4.13)$$
as the series on the right hand side is uniformly convergent.  Now, by 
using (4.13) and the formula for $a_{1}(t)$ in (4.4), we find
$$(1-e^{2}_{1}(t))p_{1}(t)\left(a_{1}(t) - \frac{1}{\lambda_{k}}\right)
= \frac{2\lambda_{k} - p_{1}(t)(1-e^{2}_{1}(t))}{\lambda_{k}}
\sim\, 2 \quad \hbox{as} \,\, t\to \infty.\eqno(4.14)$$
As a result, it follows from the recurrence relation (4.5) for $j=1$ and 
(4.11) that
$$\lim_{t\to\infty} e_{1}(t)\frac{\phi_{k}(2,t)}{\sqrt{p_{2}(t)}} =0.\eqno(4.15)$$ 
Hence 
$$\lim_{t\to\infty} e_{1}(t) =\lim_{t\to\infty} \left(2 \sum_{k=1}^{\infty} \lambda_{k}
  \frac{\phi_{k}(1,t)}{\sqrt{p_{1}(t)}}e_{1}(t) \frac{\phi_{k}(2,t)}
 {\sqrt{p_{2}(t)}}\right)^{1\over 2} =0\eqno(4.16)$$
by (4.11), (4.15)
and so we can make the following improvement to (4.14):
$$p_{1}(t)\left(a_{1}(t) - \frac{1}{\lambda_{k}}\right)\sim\, 2 \quad \hbox{as} \,\, t\to \infty.\eqno(4.17)$$
Next, we claim that
$$p_{2}(t)b_{1}(t) \frac{\phi_{k}(1,t)}{\phi_{k}(2,t)} \sim\, 2 e^{2}_{1}(t)\quad 
\hbox{as} \,\,  t\to \infty.\eqno(4.18)$$
To see that this is true, use the recurrence relation (4.6) for 
$j=1$ and the formula for $b_{1}(t)$ in (4.4), we find that
$$p_{2}(t)b_{1}(t) \frac{\phi_{k}(1,t)}{\phi_{k}(2,t)} = 
\frac{4 e^{2}_{1}(t)}{(1-e^{2}_{1}(t))^{2}} \frac{1}{p_{1}(t)(a_{1}(t)-{1\over \lambda_{k}})}.\eqno(4.19)$$
Therefore the claim follows from (4.16) and (4.17).  Note
that in particular, the relation in (4.18) implies that 
$$\phi_{k}(2,t) >0 \quad \hbox{for}\,\, t\,\, \hbox{sufficiently large}.
\eqno(4.20)$$

We next use the recurrence relation (4.6) for $j=1$, (4.9) and
(4.17) to obtain
$$\eqalign{
\frac{\phi_{r}(2,t)}{\phi_{s}(2,t)} = & \frac{p_{1}(t)(a_{1}(t) -{1\over \lambda_{r}})\phi_{r}(1,t)} {p_{1}(t)(a_{1}(t) -{1\over \lambda_{s}})\phi_{s}(1,t)}\cr
\sim & e^{-{1\over 2}(\lambda_r -\lambda_s)t}
  \frac{\phi_{r}(1,0)}{\phi_{s}(1,0)} \quad \hbox{as}\,\, t\to \infty.\cr}
\eqno(4.21)$$
From this, it follows that
$$\eqalign{
\frac{\phi_{k}^{2}(2,t)}{p_{2}(t)}
  =&\,\frac{\phi_{k}^{2}(2,t)}{2\sum_{j=1}^{\infty} \lambda_{j}\phi^{2}_{j}(2,t)}\cr
  <&\,\frac{1}{2\lambda_{k+1}} \frac{\phi_{k}^{2}(2,t)}{\phi_{k+1}^{2}(2,t)}\cr
  \sim &\, c_{k1}\,e^{-(\lambda_{k}-\lambda_{k+1})\,t}\quad \hbox{as}\,\, t\to \infty.
   \cr}\eqno(4.22)$$
Consequently, from (4.22) and (4.20), we have
$$\lim_{t\to\infty} \frac{\phi_{k}(2,t)}{\sqrt{p_{2}(t)}}=0.\eqno(4.23 )$$ 
Therefore, as $\lim_{t\to\infty} p_{2}(t) = \alpha^{-}_{2}<\infty,$ we
also obtain from (4.22) that
$$\lim_{t\to \infty} \phi_{k}(2,t) =0\eqno(4.24)$$
and so
$$\lim_{t\to\infty} p_{2}(t)=\sum_{j=1}^{\infty} \lambda_{j} \phi^{2}_{j}(2,t)=0.
\eqno(4.25)$$
Hence it follows from the formula for $a_{2}(t)$ together with
(4.16) and (4.25) that
$$\eqalign{&(1-e^{2}_{2}(t))p_{2}(t)\left(a_{2}(t) - \frac{1}{\lambda_{k}}\right)\cr
= &\frac{2\lambda_{k}(1-e^{2}_{1}(t)e^{2}_{2}(t)) - p_{2}(t)(1-e^{2}_{1}(t))
   (1-e^{2}_{2}(t))}{\lambda_{k}(1-e^{2}_{1}(t))}\cr
\sim &\, 2 \quad \hbox{as} \,\, t\to \infty.\cr}\eqno(4.26)$$
Using the recurrence relation in (4.5) for $j=2$, (4.11), (4.16), (4.23) and
(4.26), we now conclude that
$$\lim_{t\to\infty} e_{2}(t)\frac{\phi_{k}(3,t)}{\sqrt{p_{3}(t)}} =0.\eqno(4.27)$$
As a consequence of (4.27) and (4.23), we find
$$\lim_{t\to\infty} e_{2}(t) =\lim_{t\to\infty} \left(2 \sum_{k=1}^{\infty} \lambda_{k}
  \frac{\phi_{k}(2,t)}{\sqrt{p_{2}(t)}}e_{2}(t) \frac{\phi_{k}(3,t)}
  {\sqrt{p_{3}(t)}}\right)^{1\over 2} =0\eqno(4.28)$$ 
and so we can improve (4.26) to
$$p_{2}(t)\left(a_{2}(t) - \frac{1}{\lambda_{k}}\right)\sim\, 2 \quad \hbox{as} \,\, t\to \infty.\eqno(4.29)$$
We next show
$$p_{3}(t)b_{2}(t) \frac{\phi_{k}(2,t)}{\phi_{k}(3,t)} \sim\, 2 e^{2}_{2}(t)\quad 
\hbox{as} \,\,  t\to \infty.\eqno(4.30)$$
To establish this, we make use of the recurrence relation (4.6) for
$j=2$ and the formula for $b_{2}(t)$ in (4.4), this yields
$$\eqalign{
 p_{3}(t)b_{2}(t) \frac{\phi_{k}(2,t)}{\phi_{k}(3,t)}=\,&  
\frac{ p_{3}(t)b^{2}_{2}(t)}{(a_{2}(t) -{1\over \lambda_{k}}) -
     b_{1}(t)\frac{\phi_{k}(1,t)}{\phi_{k}(2,t)}}\cr
=\, & \frac{4 e^{2}_{2}(t)}{(1-e^{2}_{2}(t))^{2}} \frac{1}{p_{2}(t)(a_{2}(t)-
      {1\over \lambda_{k}})-p_{2}(t)b_{1}(t)\frac{\phi_{k}(1,t)}{\phi_{k}(2,t)}}.
\cr}\eqno(4.31)$$
from which we obtain the assertion by using  (4.18), (4.28) and
(4.29).  Note that if we combine (4.30) with (4.20), we conclude that
$$\phi_{k}(3,t) >0 \quad \hbox{for}\,\, t\,\, \hbox{sufficiently large}.
\eqno(4.32)$$

Now let $n\geq 3$ and assume by induction that the following sequence of
assertions holds for $j\leq n-1$ for each $k, r, s\in \Bbb N$:
$$\eqalign{
(1)_{j} \quad&\frac{\phi_{r}(j,t)}{\phi_{s}(j,t)}\sim e^{-{1\over 2}(\lambda_r -\lambda_s)t}
  \frac{\phi_{r}(1,0)}{\phi_{s}(1,0)} \quad \hbox{as}\,\, t\to \infty,\cr
(2)_{j} \quad& \frac{\phi_{k}^{2}(j,t)}{p_{j}(t)} < \frac{1}{2\lambda_{k+1}} 
\frac{\phi_{k}^{2}(j,t)}{\phi_{k+1}^{2}(j,t)}\sim
c_{k1}\,e^{-(\lambda_{k}-\lambda_{k+1})\,t}\quad \hbox{as}\,\, t\to \infty,\cr
(3)_{j} \quad&\lim_{t\to\infty} \frac{\phi_{k}(j,t)}{\sqrt{p_{j}(t)}}=0,\cr 
(4)_{j} \quad&\lim_{t\to \infty} \phi_{k}(j,t) =0,\cr
(5)_{j} \quad&\lim_{t\to\infty} p_{j}(t)=0,\cr
(6)_{j} \quad&(1-e^{2}_{j}(t))p_{j}(t)\left(a_{j}(t) - \frac{1}{\lambda_{k}}\right)
\sim\, 2 \quad \hbox{as} \,\, t\to \infty,\cr
(7)_{j} \quad&\lim_{t\to\infty} e_{j}(t)\frac{\phi_{k}(j+1,t)}{\sqrt{p_{j+1}(t)}} =0,
\cr
(8)_{j} \quad&\lim_{t\to\infty} e_{j}(t) = 0,\cr
(9)_{j} \quad&p_{j}(t)\left(a_{j}(t) - \frac{1}{\lambda_{k}}\right)\sim\, 2 \quad \hbox{as} \,\, t\to \infty,\cr
(10)_{j} \quad&p_{j+1}(t)b_{j}(t) \frac{\phi_{k}(j,t)}{\phi_{k}(j+1,t)} \sim\, 2 
e^{2}_{j}(t)\quad 
\hbox{as} \,\,  t\to \infty.\cr}\eqno(4.33)$$

We shall prove the sequence of assertions holds for $j=n.$  We begin by
invoking the recurrence relation (4.6) for $j=n-1$, this gives
$$\eqalign{
\frac{\phi_{r}(n,t)}{\phi_{s}(n,t)}=\,&
\frac{\left(a_{n-1}(t) -{1\over \lambda_{r}}\right)\phi_{r}(n-1,t) - 
b_{n-2}(t)\phi_{r}(n-2,t)}
{\left(a_{n-1}(t) -{1\over \lambda_{s}}\right)\phi_{s}(n-1,t) - 
b_{n-2}(t)\phi_{s}(n-2,t)}
\cr
=\,&\frac{\phi_{r}(n-1,t)}{\phi_{s}(n-1,t)}
\frac{p_{n-1}(t)\left(a_{n-1}(t) - {1\over \lambda_{r}}\right) - p_{n-1}(t) b_{n-2}(t)
\frac{\phi_{r}(n-2,t)}{\phi_{r}(n-1,t)}}
{p_{n-1}(t)\left(a_{n-1}(t) -{1\over \lambda_{s}}\right) - p_{n-1}(t) b_{n-2}(t)
\frac{\phi_{s}(n-2,t)}{\phi_{s}(n-1,t)}}\cr
\sim\,&e^{-{1\over 2}(\lambda_r -\lambda_s)t}
  \frac{\phi_{r}(1,0)}{\phi_{s}(1,0)} \quad \hbox{as}\,\, t\to \infty\cr}\eqno(4.34)
$$
by the induction assumptions $(1)_{n-1}$, $(9)_{n-1}$,$(10)_{n-2}$ and 
$(8)_{n-2}$.  
Thus as before, we have
$$\frac{\phi_{k}^{2}(n,t)}{p_{n}(t)}  < \frac{1}{2\lambda_{k+1}} 
\frac{\phi_{k}^{2}(n,t)}{\phi_{k+1}^{2}(n,t)}\sim
c_{k1}\,e^{-(\lambda_{k}-\lambda_{k+1})\,t}\quad \hbox{as}\,\, t\to \infty\eqno(4.35)$$
from which it follows that
$$\lim_{t\to\infty} \frac{\phi_{k}(n,t)}{\sqrt{p_{n}(t)}}=0\eqno(4.36)$$
since $\phi_{k}(n,t) >0$ for $t$ sufficiently large by the induction
assumptions.  Consequently,
$$\lim_{t\to \infty} \phi_{k}(n,t) =0\eqno(4.37)$$
and so
$$\lim_{t\to\infty} p_{n}(t)=\lim_{t\to\infty}2 \sum_{j=1}^{\infty} 
\lambda_{j}\phi^{2}_{j}(n,t) =0.\eqno(4.38)$$
To establish $(6)_{n}$, we make use of the formula for $a_{n}(t)$ in (4.4),
the induction assumptions and (4.38), thus 
$$\eqalign{&(1-e^{2}_{n}(t))p_{n}(t)\left(a_{n}(t) - \frac{1}{\lambda_{k}}\right)\cr
= &\frac{2\lambda_{k}(1-e^{2}_{n-1}(t)e^{2}_{n}(t)) - p_{n}(t)(1-e^{2}_{n-1}(t))
   (1-e^{2}_{n}(t))}{\lambda_{k}(1-e^{2}_{n-1}(t))}\cr
\sim &\, 2 \quad \hbox{as} \,\, t\to \infty.\cr}\eqno(4.39)$$
Therefore, on using the recurrence relation (4.5) for $j=n$,
the induction assumptions $(3)_{n-1},$ $(8)_{n-1}$ together
with (4.36) and (4.39), we obtain
$$\lim_{t\to\infty} e_{n}(t)\frac{\phi_{k}(n+1,t)}{\sqrt{p_{n+1}(t)}} =0.\eqno(4.40)$$
Hence$$\lim_{t\to\infty} e_{n}(t) =\lim_{t\to\infty} \left(2 \sum_{k=1}^{\infty} 
\lambda_{k}\frac{\phi_{k}(n,t)}{\sqrt{p_{n}(t)}}e_{n}(t) \frac{\phi_{k}(n+1,t)}
{\sqrt{p_{n+1}(t)}}
  \right)^{1\over 2} =0\eqno(4.41)$$
by (4.36) and (4.40).  Using this result in (4.39), we obtain
$(9)_{n}.$   Finally, the assertion $(10)_{n}$ follows from
$$\eqalign{&
 p_{n+1}(t)b_{n}(t) \frac{\phi_{k}(n,t)}{\phi_{k}(n+1,t)}\cr 
=\,&  
\frac{ p_{n+1}(t)b^{2}_{n}(t)}{(a_{n}(t) -{1\over \lambda_{k}}) -
     b_{n-1}(t)\frac{\phi_{k}(n-1,t)}{\phi_{k}(n,t)}}\cr
=\, & \frac{4 e^{2}_{n}(t)}{(1-e^{2}_{n}(t))^{2}} \frac{1}{p_{n}(t)(a_{n}(t)-
      {1\over \lambda_{k}})-p_{n}(t)b_{n-1}(t)\frac{\phi_{k}(n-1,t)}{\phi_{k}(n,t)}}\cr}\eqno(4.42)$$
upon using $(9)_{n}$, the induction assumptions $(8)_{n-1}$,
$(10)_{n-1}$ and (4.41).
This completes the proof of the sequence of assertions $(1)_{j}$-$(10)_{j}$ by
induction.  Hence we have established parts (a) and (b) of
the theorem as
the relation $\lim_{t\to\infty} e_{j}(t) =0$ is 
equivalent to $\lim_{t\to\infty} (q_{j+1}(t) -q_{j}(t)) = \infty.$
To prove part (c), we begin with the assertion for $j=1.$
For this case, note that from the equation for $p_1(t)$, we
have
$$p_1(t) =p_1(0)\, exp\left(-{1\over 2}\int_{0}^{t} \sum_{k\neq 1}
  e^{-(q_{k}(s) -q_1(s))} p_k(s)\,ds\right).\eqno(4.43)$$
Since $\lim_{t\to\infty} p_1(t) =0,$ we conclude from (4.43) that
$$\int_{0}^{t} \sum_{k\neq 1} e^{-(q_{k}(s) -q_1(s))} p_k(s)\,ds\to\infty\quad
  \hbox{as}\,\,\, t\to\infty.\eqno(4.44)$$
Meanwhile, from the equation for $q_1(t)$ in (3.6), we find
$$\eqalign{
q_1(t) =\,& q_1(0) + {1\over 2}\int_{0}^{t} \sum_{k=1}^{\infty}
            e^{-|q_{k}(s) -q_1(s)|} p_k(s)\,ds\cr
       >\,& q_1(0) + {1\over 2}\int_{0}^{t} \sum_{k\neq1}
            e^{-(q_{k}(s) -q_1(s))} p_k(s)\,ds,\quad t>0.\cr}\eqno(4.45)$$
Therefore, on taking the limit as $t$ tends to infinity in (4.45)
and making use of (4.44), we conclude that $q_1(t)\to\infty$ as
$t\to \infty.$  To show that $q_j(t)\to\infty$ as $t\to\infty$
for $j>1,$ note that $\dot q_{j-1}(t) >0$ by Proposition 3.4 (b).
Since $(q(t),p(t))\in {\Cal S}_{-}$, it follows from this
property that
$$0 < q_j(t)-q_{j-1}(t) < q_j(t)-q_{j-1}(0) \quad \hbox{for all}\,\, t>0.
\eqno(4.46)$$
Hence the assertion follows from (4.46) as we have
$\lim_{t\to\infty} (q_j(t) -q_{j-1}(t)) = \infty.$
To establish part (d), note that by parts (a) and (b) above, and
the equation for $q_j(t)$ in (3.6), we have
$\lim_{t\to\infty} \dot q_j(t) =0.$  Since $q_j(t)\to \infty$ as $t\to\infty$,
it follows by the L'H\^opital's rule that
$$\lim_{t\to\infty} \frac{q_j(t)}{t} = \lim_{t\to\infty} \dot q_j(t) =0.\eqno(4.47)$$
Hence $q_j(t) = o(t)$ as $t\to \infty,$ as asserted.

\pf
\enddemo

\medskip

\proclaim
{Corollary 4.4} If 
$u_{0}(x) = {1\over 2}\sum_{j=1}^{\infty} e^{-|x-q^{0}_{j}|} p^{0}_j$
where $(q^0,p^0)\in {\Cal S}_{-}$, then the solution $u(x,t)$
of the CH equation (3.1) with $u(x,0) = u_{0}(x)$ is such that
$$u(x,t) \simeq 0 \quad \hbox{as}\,\,\,t\to \infty.\eqno(4.48)$$
\endproclaim

\noindent{\bf Remark 4.5} (a)  Theorem 4.3 (a) can also be proved
using the method in Section 5 below.  
\newline
(b) From the fact that $\overline{Ran \bl(q,p)} = {\Cal H}$, it
is straightforward to show that the equation
$\dot \bl(q,p) = {1\over 2} [\,\bl(q,p), \Pi_{\frak k} \bl(q,p)\,]$
implies
$$\dot \bj(q,p) = {1\over 2} [\,\bj(q,p), \Pi_{\frak k} \bl(q,p)\,].$$
However, as $\bl(q,p)$ is not invertible, it is not possible to
express this equation in terms of $\bj(q,p)$ alone.
\newline
(c) From the definition of $a_j(t)$ in (4.4) and the proof of Theorem 4.3
above, we see that $a_j(t)\to \infty$ as $t\to\infty.$

\bigskip
\bigskip

\subhead
5. \ Long time behaviour in ${\Cal S}_{+}$ and other sectors
\endsubhead

\bigskip

Let $(q(t), p(t))$ be the solution of (3.6) with 
$(q(0), p(0))= (q^0,p^0)\in {\Cal S}_{+}.$ Then
$\bl(t)=\bl(q(t),p(t))$ evolves under the (+) Toda flow with
initial condition $\bl(0) = \bl(q^0,p^0)$.  As in Section 4, 
we order the eigenvalues of $\bl(q^0, p^0)$ in such a way that
$0 < \cdots < \lambda_3 < \lambda_2 < \lambda_1.$  Also,
we let $\phi_{k}(t)$ be the normalized eigenvector
of $\bl(t)$ corresponding to $\lambda_k$ with $\phi_{k}(1,t)>0,$
$k=1, 2,\cdots.$

Denote by $\wedge^{k} {\Cal H}$ the $k$-th exterior power of ${\Cal H},$
$k\geq 1.$ (See \c{LS} and \c{RS2} for more details.)
Then the operator $\bl(t)$ gives rise to the induced derivations
$$(\bl(t))_{k} \equiv \sum_{r=0}^{k-1} \bi^{r}\otimes \bl(t)\otimes \bi^{k-1-r}:
\wedge^{k} {\Cal H}\longrightarrow \wedge^{k}{\Cal H},\quad k\geq 1\eqno(5.1)$$
where $\bi^{r}$ and $\bi^{k-1-r}$ are the identity operators on 
$\wedge^{r} {\Cal H}$ and $\wedge^{k-1-r} {\Cal H}$ respectively.   Since 
ker\,$\bl(t)=\{0\},$ $\{\phi_{k}(t)\}_{k=1}^{\infty}$ is an orthonormal
basis of ${\Cal H}$.   Thus 
$$\{\phi_{i_1}(t)\wedge\cdots \wedge\phi_{i_k}(t)\mid 1\leq i_1<\cdots <i_k\}
\eqno(5.2)$$
is an orthonormal basis of $\wedge^{k}{\Cal H}$
with respect to the natural inner product on $\wedge^{k}{\Cal H}$
defined by
$$(\xi_1\wedge\cdots\wedge\xi_k, \eta_1\wedge\cdots\wedge\eta_k)
  = det ((\xi_i,\eta_j)).\eqno(5.3)$$
Moreover, the elements $\phi_{i_1}(t)\wedge\cdots \wedge\phi_{i_k}(t)$
($i_1<\cdots<i_k$)
are eigenvectors of $(\bl(t))_{k}$ as we have
$$(\bl(t))_{k}(\phi_{i_1}(t)\wedge\cdots \wedge\phi_{i_k}(t))
=(\lambda_{i_1}+\cdots + \lambda_{i_k})(\phi_{i_1}(t)\wedge\cdots\wedge\phi_{i_k}(t)).\eqno(5.4)$$

\proclaim
{Lemma 5.1} For any increasing sequence $1\leq i_1<\cdots <i_k$ of
numbers from $\Bbb N,$
$$\eqalign{
&(e_1\wedge\cdots\wedge e_k, \phi_{i_1}(t)\wedge\cdots\wedge \phi_{i_k}(t))^{2}\cr
=\,\,&\frac{e^{(\lambda_{i_1}+\cdots + \lambda_{i_k})t}(e_1\wedge\cdots\wedge e_k, 
\phi_{i_1}(0)\wedge\cdots \wedge\phi_{i_k}(0))^{2}}
{\sum_{1\leq j_1<\cdots<j_k}
 e^{(\lambda_{j_1}+\cdots + \lambda_{j_k})t}(e_1\wedge\cdots\wedge e_k, \phi_{j_1}(0)
 \wedge\cdots \wedge\phi_{j_k}(0))^{2}}.\cr}\eqno(5.5)$$
Hence 
$$\lim_{t\to\infty}(e_1\wedge\cdots\wedge e_k, \phi_{i_1}(t)\wedge\cdots\wedge 
\phi_{i_k}(t))^{2}=\cases 1, &\text{for $(i_1,\cdots,i_k)=(1,\cdots,k)$}\\
                        0, &\text{otherwise}.\endcases\eqno(5.6)$$
\endproclaim

\demo
{Proof} By Theorem 2.9, if $b_{\pm}(t)$ are the solutions of the
factorization problem $e^{-{1\over 2}t\bl(0)} = b_{+}(t)b_{-}(t)^{-1}$
with $b_{+}(t)\in {\Cal K}$ and $b_{-}(t)\in {\Cal L}$, then
$\phi_{j}(t) = b_{+}(t)^{-1}\phi_{j}(0)$ for all $j.$
Therefore,
$$\eqalign{
&(e_1\wedge\cdots\wedge e_k, \phi_{j_1}(t)\wedge\cdots\wedge \phi_{j_k}(t))\cr
=\,\,&(e_1\wedge\cdots\wedge e_k, (b_{+}(t)^{-1})^{\wedge\, k}\phi_{j_1}(0)\wedge
\cdots\wedge \phi_{j_k}(0))\cr
=\,\,&(((b_{-}(t)^{-1})^{T})^{\wedge\, k}e_1\wedge\cdots\wedge e_k,
      (e^{{1\over 2}t\bl(0)})^{\wedge\, k}\phi_{j_1}(0)\wedge\cdots \wedge\phi_{j_k}(0))\cr
=\,&\frac{e^{{1\over 2}(\lambda_{j_1}+\cdots + \lambda_{j_k})\,t}(e_1\wedge\cdots\wedge e_k, 
\phi_{j_1}(0)\wedge\cdots \wedge\phi_{j_k}(0))}
{(b_{-}(t))_{11}\cdots (b_{-}(t))_{kk}}\cr}\eqno(5.7)$$
for any increasing sequence $1\leq j_1<\cdots <j_k.$
Consequently, upon substituting the above expression into
$$1\,\, = \sum_{1\leq j_1<\cdots<j_k}
(e_1\wedge\cdots\wedge e_k, \phi_{j_1}(t)
 \wedge\cdots \wedge\phi_{j_k}(t))^{2},\eqno(5.8)$$
we obtain
$$\eqalign{
&((b_{-}(t))_{11}\cdots (b_{-}(t))_{kk})^{2}\cr
=\,\, &{\sum_{1\leq j_1<\cdots<j_k}
 e^{(\lambda_{j_1}+\cdots + \lambda_{j_k})t}(e_1\wedge\cdots\wedge e_k, \phi_{j_1}(0)
 \wedge\cdots \wedge\phi_{j_k}(0))^{2}}.\cr}\eqno(5.9)$$
Hence (5.5) follows from (5.7) and (5.9).  The assertion in (5.6) is now 
seen to be a corollary of the fact that
$$\eqalign{
&\sum_{1\leq j_1<\cdots<j_k}
 e^{(\lambda_{j_1}+\cdots + \lambda_{j_k})t}(e_1\wedge\cdots\wedge e_k, \phi_{j_1}(0)
 \wedge\cdots \wedge\phi_{j_k}(0))^{2}\cr
\sim\,\,& e^{(\lambda_{1}+\cdots + \lambda_{k})t}(e_1\wedge\cdots\wedge e_k, \phi_{1}(0)
 \wedge\cdots \wedge\phi_{k}(0))^{2} \quad \hbox{as}\,\,t\to\infty.\cr}
\eqno(5.10)$$
\pf
\enddemo

\proclaim
{Theorem 5.2} Let $(q(t), p(t))$ be the solution of (3.6) with
$(q(0), p(0))= (q^0,p^0)\in {\Cal S}_{+}.$ Then 
\newline
\noindent (a)  $\lim_{t\to \infty} p_j(t) =\, 2\lambda_j$ for all $j\in \Bbb N,$
\newline
\noindent (b) $\lim_{t\to \infty} |q_{j}(t) -q_{k}(t)| =\infty$ for all
$j\neq k,$
\newline
\noindent (c) $q_j(t)\sim \lambda_j t\quad \hbox{as}\,\,\, t\to\infty$
for all $j\in \Bbb N.$
\endproclaim 

\demo
{Proof} For all $k\in \Bbb N$, 
$$\eqalign{
&\sum_{i=1}^{k} L_{ii}(t)\cr
=\,\,& (e_1\wedge\cdots\wedge e_k, (\bl(t))_{k} e_1\wedge\cdots\wedge e_k)\cr
=\,\,&\sum_{1\leq i_1<\cdots<i_k} (\lambda_{i_1} + \cdots + \lambda_{i_k})
       (e_1\wedge\cdots\wedge e_k, \phi_{i_1}(t)\wedge\cdots\wedge\phi_{i_k}(t))^{2}
\cr}\eqno(5.11)$$
where we have used (5.4) and the fact that 
$\{\phi_{i_1}(t)\wedge\cdots \phi_{i_k}(t)\mid 1\leq i_1<\cdots <i_k\}$
is an orthonormal basis of $\wedge^{k} {\Cal H}.$
Taking the limit as $t\to\infty$ in (5.11), and using Proposition 3.4 (a) and
(5.6), we find
$$\alpha^{+}_1 + \cdots \alpha^{+}_{k} = \lambda_1 + \cdots + \lambda_k.\eqno(5.12)
$$
Consequently we have 
$$\alpha^{+}_{k} = \lambda_k\eqno(5.13)$$
for all $k$ and this proves part (a).   The assertion in part (b) is
now a consequence of part (a) and Proposition 3.4 (b).  To establish
part (c), note that all the terms on the right hand side of the
equation for $\dot q_j(t)$ in (3.6) tend to zero as $t\to\infty$
except for the term ${1\over 2} p_j(t)$.  Hence 
$\dot q_j(t)\sim \lambda_j$ as $t\to\infty$ from which the
assertion follows.
\pf
\enddemo

\smallskip

\proclaim
{Corollary 5.3} If 
$u_{0}(x) = {1\over 2}\sum_{j=1}^{\infty} e^{-|x-q^{0}_{j}|} p^{0}_j$
where $(q^0,p^0)\in {\Cal S}_{+}$, then the solution $u(x,t)$
of the CH equation (3.1) with $u(x,0) = u_{0}(x)$ is such that
$$u(x,t) \simeq \sum_{j=1}^{\infty} \lambda_{j}\,e^{-|x-\lambda_{j}t|}\quad \hbox{as}
\,\,\, t\to\infty.\eqno(5.14)$$
\endproclaim
\medskip
\noindent{\bf Remark 5.4} (a) In \c{DLT1}, the authors study the
long time behaviour of the Toda flows on {\it general \/} bounded symmetric
operators in $l^{+}_{2}$.  It is in this context that the induced
derivations of the Lax operator was first introduced.  Indeed,
the quantity corresponding to the one which appears on the second line of 
(5.11) (with $\bl(t)$ replaced by the solution of the Toda flow on a 
{\it general \/} symmetric operator on $l^{+}_{2}$)
was also considered in Lemma 1, Section 4
of \c{DLT1}. The difference here is that there is no
need for us to introduce spectral measures as we can do
everything explicitly in terms of eigenvalues and eigenvectors. 
\newline
(b) The proof of Theorem 5.2 above shows that the long time
behaviour of $\bl(t)$ has the sorting property, as in the
case of the Toda flow on $N\times N$ Jacobi matrices \c{Mo}.  
Note that this is not true for the Toda flow on {\it general \/}
symmetric operators on $l^{+}_{2}$ \c{DLT1}.
\newline
(c) In their analysis of the multipeakon solutions in the
sector where $q_1 <\cdots <q_N$, the authors in \c{BSS} found that
$\lim_{t\to -\infty} p_j(t) = \lim_{t\to\infty} p_{N+1-j}(t)$ for
$j=1,\cdots, N$.   However, from our investigation in
Section 4 and in the present section, we see that there
is no analog of this relation in our case.
\medskip

Note that the behaviour in (5.14) is given in a convincing but
heuristic discussion in \c{M} with no explicit assumptions
on the ordering of the $q_j$'s.  In view of this, we ask
if there are other sectors in the phase space besides ${\Cal S_{+}}$
for which this behaviour is valid.  We give an answer to this
question as follows.  Let Aut$(\Bbb N)$ be the group of all
permutations of $\Bbb N$ and let
$$\Sigma=\{p=(p_1,p_2,\cdots)\in l^{+}_{1,2}\mid p_j >0\,\,
\hbox{for all}\,\, j\}.\eqno(5.15)$$
If $\pi\in \,\,\hbox{Aut}(\Bbb N),$ $p=(p_1,p_2,\cdots)\in l^{+}_{1,2}$
and $q=(q_1,q_2,\cdots)\in l^{+}_{\infty},$
define $\pi\cdot p = (p_{\pi(1)},p_{\pi(2)},\cdots)$
and similarly for $\pi\cdot q.$  With this
notation, we set
$$\hbox{Aut}(\Bbb N)_{\Sigma} =\{\pi\in \hbox{Aut}(\Bbb N)\mid \pi\cdot p\in
\Sigma \,\,\,\hbox{for all}\,\,p\in \Sigma \}.\eqno(5.16)$$
Clearly, $\hbox{Aut}(\Bbb N)_{\Sigma}$ contains the finite
permutations of $\Bbb N,$ it also contains those permutations
that satisfy the condition $\lim_{j\to\infty} \pi(j)/j =1,$ for
example.  For $\pi\in \hbox{Aut}(\Bbb N)_{\Sigma},$ consider
the sector 
$${\Cal S}_{+}(\pi) = \lbrace (q,p)\in l^{+}_{\infty}\oplus l^{+}_{1,2} \mid
   q_{\pi(1)} > q_{\pi(2)}> \cdots, p_j > 0 \,\, \hbox{for all}\,\, j \rbrace.
\eqno(5.17)$$
Note that for  $(q,p)\in {\Cal S}_{+}(\pi)$,
we can rewrite the equations of motion (3.6) in the form
$$\eqalign{
&\dot{q}_{\pi(j)} = \frac{1}{2} \sum_{k=1}^{\infty} e^{-|q_{\pi(j)} -q_{\pi(k)}|} p_{\pi(k)}\cr
&\dot{p}_{\pi(j)} = \frac{1}{2} p_{\pi(j)}\sum_{k=1}^{\infty} sgn(q_{\pi(j)} -q_{\pi(k)})
   e^{-|q_{\pi(j)} -q_{\pi(k)}|} p_{\pi(k)},\cr
&\phantom{abcd} = \frac{1}{2} p_{\pi(j)}\sum_{k=1}^{\infty} sgn(k-j)
   e^{-|q_{\pi(j)} -q_{\pi(k)}|} p_{\pi(k)},\quad j\in \Bbb N\cr}\eqno(5.18)$$
as the series on the right hand side of (3.6) are 
absolutely convergent and rearrangements of such series have the same sum.
Therefore, when we compare (5.18) with (3.9), we see that the calculation
which leads to Proposition 3.2 (d) also gives
the Lax pair for (5.18). Namely,
$$\dot \bl_{\pi}(q,p) = 
{1\over 2} [\Pi_{\frak k} \bl_{\pi}(q,p)\, \bl_{\pi}(q,p) \,] \eqno(5.19)$$
where $\bl_{\pi}(q,p) = \left({1\over 2} e^{-{1\over 2}|q_{\pi(i)}-q_{\pi(j)}|}
\sqrt{p_{\pi(i)}p_{\pi(j)}}\right)_{i,j=1}^{\infty}$ is a positive, semiseparable
trace class operator in $\l^{+}_{2}$ with simple eigenvalues
$0 < \cdots < \lambda_3 < \lambda_2 < \lambda_1.$  Hence 
Proposition 5.2 can be applied to 
$(\pi\cdot q(t), \pi\cdot p(t))$
and we conclude that for all $j\in \Bbb N$,
$$\lim_{t\to \infty} p_j(t) =\, 2\lambda_{\pi^{-1}(j)},\quad
q_j(t)\sim \lambda_{\pi^{-1}(j)} t\quad \hbox{as}\,\,\, t\to\infty.\eqno(5.20)
$$
Consequently the long time asymptotics for $u(x,t)$ is given by (5.14), 
as before. 
 
Clearly, similar consideration and Theorem 4.3 can be applied
to obtain the long time behaviour in sectors of the form
$${\Cal S}_{-}(\pi) = \lbrace (q,p)\in l^{+}_{\infty}\oplus l^{+}_{1,2} \mid
   q_{\pi(1)} < q_{\pi(2)} < \cdots, p_j > 0 \,\, \hbox{for all}\,\, j \rbrace
\eqno(5.21)$$
for $\pi\in \hbox{Aut}(\Bbb N)_{\Sigma}$ with the following result:
$$\lim_{t\to\infty} p_j(t) =0,\quad q_j(t)\to\infty, \quad q_j(t) =o(t)\quad
\hbox{as}\,\,\, t\to\infty\eqno(5.22)$$
for all $j\in \Bbb N.$  Hence (4.48) remains valid for
$(q^0,p^0)\in {\Cal S}_{-}(\pi).$

\medskip
\noindent{\bf Acknowledgments.}  It is a pleasure to thank
Percy Deift and Irina Nenciu for discussions when a
sketch of the proof of Theorem 4.3 was presented to them
while the first draft of this work was in its final stage
of preparation. The author would also like to thank
Henry McKean whose paper \c{M} has been a source of
inspiration for section 5 of the present work.


\bigskip
\bigskip


\newpage   

\Refs
\widestnumber\key{RSTS}

\ref\key{BSS} 
\by Beals, R., Sattinger, D. and Szmigielski, J.
\paper Multipeakons and the classical moment problem
\jour Adv. Math.\vol 154\yr 2000\pages 229-257
\endref

\ref\key{CH}
\by Camassa, R. and Holm, D.
\paper An integrable shallow water equation with peaked solitons
\jour Phys. Rev. Lett.\vol 71\yr 1993\pages 1661-1664
\endref

\ref\key{CHH}
\by Camassa, R., Holm, D. and Hyman, J.M.
\paper A new integrable shallow water equation
\jour Adv. Appl. Mech.\vol 31\yr 1994\pages 1-33
\endref

\ref\key{CF}
\by Calogero, F. and Francoise, J.-P.
\paper A completely integrable Hamiltonian system
\jour J. Math. Phys.\vol 37\yr 1996\pages 2863-2871
\endref

\ref\key{DLT1}
\by Deift, P., Li, L.-C. and Tomei, C.
\paper Toda flows with infinitely many variables
\jour J. Funct. Analy.\vol 64\yr 1985\pages 358-402
\endref

\ref\key{DLT2}
\by Deift, P., Li, L.-C. and Tomei, C.
\paper Matrix factorizations and integrable systems
\jour Comm. Pure Appl. Math.\vol 42\yr 1989\pages 443-521
\endref

\ref\key{GK}
\by Gantmacher, F. and Krein, M.
\book Oscillation matrices and kernels and small vibrations
of mechanical systems. Revised edition
\publ AMS Chelsea Publishing \publaddr  Providence, R.I.\yr 2002
\endref

\ref\key{L1}
\by Li, L.-C.
\paper Long time behaviour of an infinite particle system
\jour Commun. Math. Phys.\vol 110\yr 1987\pages 617-623
\endref

\ref\key{L2}
\by Li, L.-C.
\paper Factorization problem on the Hilbert-Schmidt group and the
Camassa-Holm equation
\jour Comm. Pure Appl. Math., to appear
\endref

\ref\key{LS}
\by Loomis, L. and Sternberg, S.
\book Advanced calculus
\publ Addision Wesley \publaddr Reading, Massachusetts\yr 1968
\endref

\ref\key{M}
\by McKean, H.
\paper Fredholm determinants and the Camassa Holm hierarchy
\jour Comm. Pure Appl. Math.\vol 56 \yr 2003\pages 638-680
\endref

\ref\key{Mo}  
\by Moser, J.
\paper Finitely  many mass points on the line under the
influence of an exponential potential-an integrable system
\inbook Dynamical systems, theory and applications
\bookinfo Lecture notes in Physics, vol. 38
\publ Springer-Verlag \publaddr Berlin\yr 1975\pages 467-497
\endref

\ref\key{RB}
\by Ragnisco, O. and Bruschi, M.
\paper Peakons, r-matrix and Toda lattice
\jour Phys. A\vol 228\yr 1996\pages 150-159
\endref

\ref\key{RS1}
\by Reed, M. and Simon, B.
\book Methods of mathematical physics I: functional analysis
\publ Academic Press \publaddr  New York-London\yr 1972
\endref

\ref\key{RS2}
\by Reed, M. and Simon, B.
\book Methods of mathematical physics IV: analysis of operators
\publ Academic Press  \publaddr New York-London\yr 1972
\endref

\ref\key{RSTS}
\by Reyman, A. and Semenov-Tian-Shansky, M.
\paper Group-theoretical methods in the theory of finite-dimensional
integrable systems 
\inbook Dynamical Systems VII, Encyclopaedia of Mathematical Sciences,
\vol 16
\eds V.I. Arnold and S.P. Novikov
\publ Springer-Verlag \yr 1994\pages 116-225
\endref

\ref\key{STS}
\by Semenov-Tian-Shansky, M.
\paper Classical r-matrices, Lax equations, Poisson Lie groups
and dressing transformations
\inbook Field theory, quantum gravity and strings II (Meudon/Paris, 1985/1986)
\bookinfo Lecture Notes in Physics
\vol 280
\publ Springer-Verlag\yr 1987\pages 174-214
\endref

\ref\key{Sm}
\by Smithies, F.
\book Integral equations
\bookinfo Cambridge tracts in mathematics and mathematical
physics, no. 49
\publ Cambridge University Press \yr 1958
\endref

\ref\key{S}
\by Stieltjes, T. J.
\paper Sur la r\'eduction en fraction continue d'une s\'erie proc\'edant
suivant les puissances descendantes d'une variable
\jour Ann. Fac. Sci. Toulouse Math.\vol 3\yr 1889\pages H1-H17
\endref 





\endRefs
\enddocument